\documentclass{aa}  

\usepackage{graphicx}	
\usepackage{amsmath}	
\usepackage{threeparttable}

\begin{document}

   \title{Comprehensive view on a $z\sim6.5$ radio-loud QSO: from the radio to the optical/NIR to the X-ray band}


   \author{L. Ighina
          \inst{1,2,3}
          \and A. Caccianiga\inst{1} \and A. Moretti\inst{1}
          \and J. W. Broderick\inst{3,4,5} \and J. K. Leung\inst{6,7,8,9,10} 
          \and S. Paterson\inst{3} \and F. Rigamonti\inst{1,2,11} \and N. Seymour\inst{3}
          \and S. Belladitta\inst{12,13} \and G. Drouart\inst{3} \and T. J. Galvin\inst{14,3} \and N. Hurley-Walker\inst{3}
            }

   \institute{
    INAF, Osservatorio Astronomico di Brera, via Brera 28, 20121, Milano, Italy\\
    \email{luca.ighina@inaf.it}
    \and
    DiSAT, Universit\`a degli Studi dell'Insubria, via Valleggio 11, 22100 Como, Italy
    \and
    International Centre for Radio Astronomy Research, Curtin University, 1 Turner Avenue, Bentley, WA, 6102, Australia
    \and 
    SKA Observatory, Science Operations Centre, CSIRO ARRC, 26 Dick Perry Avenue, Kensington, WA 6151, Australia 
    \and 
    CSIRO Space \& Astronomy, PO Box 1130, Bentley, WA 6102, Australia
    \and 
    Sydney Institute for Astronomy, School of Physics, University of Sydney, NSW 2006, Australia
    \and
    CSIRO Space and Astronomy, PO Box 76, Epping, NSW, 1710, Australia
    \and 
    ARC Centre of Excellence for Gravitational Wave Discovery (OzGrav), Hawthorn, VIC 3122, Australia 
    \and
    David A. Dunlap Department of Astronomy and Astrophysics, University of Toronto, 50 St. George St., Toronto, Ontario, M5S3H4, Canada
    \and
    Racah Institute of Physics, The Hebrew University of Jerusalem, Jerusalem, 91904, Israel
    \and
    INFN, Sezione di Milano-Bicocca, Piazza della Scienza 3, I-20126 Milano, Italy
    \and
    Max Planck Institut für Astronomie, Königstuhl 17, D-69117, Heidelberg, Germany
    \and
    INAF - Osservatorio di Astrofisica e Scienza dello Spazio (OAS), Via Gobetti 93/3, 40129 Bologna - Italy
    \and
    ATNF, CSIRO Space \& Astronomy, PO Box 1130, Bentley, WA 6102, Australia
    }

   \date{Received September 15, 1996; accepted March 16, 1997}

 \abstract{We present a multi-wavelength analysis, from the radio to the X-ray band, of the redshift $z=6.44$ VIK~J2318$-$31 radio-loud (RL) quasi stellar object (QSO), one of the most distant currently known in this class. The work is based on newly obtained (uGMRT, ATCA, Chandra) as well as archival (GNIRS and X-Shooter) dedicated observations that have not been published yet. Based on the observed X-ray and radio emission, its relativistic jets are likely young and misaligned from our line of sight. Moreover, we can confirm, with simultaneous observations, the presence of a turnover in the radio spectrum at $\nu_{\rm peak} \sim 650$~MHz which is unlikely to be associated with self-synchrotron absorption. From the NIR spectrum we derived the mass of the central black hole, M$_{\rm BH}=8.1^{+6.8}_{-5.6} \times 10^8 {\rm M_{\odot}}$, and the Eddington ratio, $\lambda_{\rm EDD} = 0.8^{+0.8}_{-0.6}$, using broad emission lines as well as an accretion disc model fit to the continuum emission. 
 Given the high accretion rate, the presence of a $\sim$8$\times$10$^8$~M$_\odot$ black hole at $z=6.44$ can be explained by a seed black hole ($\sim$10$^{4}$M$_\odot$) that formed at $z\sim25$, assuming a radiative efficiency $\eta_{\rm d}\sim0.1$. However, by assuming $\eta_{\rm d}\sim0.3$, as expected for jetted systems, the mass observed would challenge current theoretical models of black hole formation.
}
 
 \titlerunning{Comprehensive view on a $z\sim6.5$ RL QSO}
   \maketitle
%

\section{Introduction}

High-redshift quasi-stellar objects (QSOs) are among the most important tools to investigate the properties of supermassive black holes (SMBHs) in the initial stages after their formation \citep[e.g.][]{Volonteri2021}. In particular, the observation of very massive BHs already at $z\sim6-7$, that is, 900--750~Myr after the Big Bang, has challenged our current understanding of BH formation and growth \citep[e.g.][]{Pacucci2017}. Indeed, in order to produce $\sim$10$^{8-9}$M$_\odot$ BHs in such a short period of time, they must originate from already massive seed BHs through the direct collapse of low-metallicity gas clouds ($\sim$10$^{5-6}$M$_\odot$; e.g. \citealt{Oh2002, Ferrara2014,Maio2019}) or through stellar-dynamical interactions in dense primordial star clusters ($\sim$10$^{3-4}$M$_\odot$; e.g. \citealt{Devecchi2009,Sakurai2017}). Otherwise, in the case of a relatively low-mass seed BH produced by, for example, the collapse of Population III stars ($\sim$10$^{2-3}$M$_\odot$; e.g. \citealt{Banik2019}), super-Eddington accretion must occur in the first phases of the SMBH evolution \citep[e.g.][]{Pezzulli2016,Takeo2019}.
Even more challenging to explain is the presence of very massive BHs hosted in radio-loud (RL)\footnote{From the observational point of view, a source is defined radio loud if the rest-frame flux density ratio $S_\mathrm{5GHz}$/$S_\mathrm{4400\text{\normalfont\AA}} >10$ \citep{Kellerman1989}.} QSOs at $z>6$. These systems are able to expel part of the accretion matter in the form of two bipolar relativistic jets (e.g. \citealt{Padovani2017}), which can extend up to mega-parsec scales \citep[e.g.][]{Bagchi2014,Mahato2022}. Indeed, the presence of relativistic jets is usually associated with highly spinning BHs \citep[e.g.][]{Blandford1977, Blandford2019}, which have a larger radiative efficiency ($\eta_{\rm d}\sim0.3$; e.g., \citealt{Thorne1974}) compared to radio-quiet (RQ; i.e., without relativistic jets; $\eta_{\rm d}\sim0.1$) QSOs. This would imply that either RL QSOs originate from the most massive seed BHs or that they must be accreting at a super-Eddington rate. In order to constrain the effect that relativistic jets have on the accretion and evolution of the SMBH that host them, statistical samples of high-$z$ RL QSOs with multiwavelength information are needed. While optical and infrared (IR) observations are crucial to investigate the accretion process and to constrain the properties of the central SMBH \citep[e.g.][]{Shen2019,Farina2022,Mazzucchelli2023}, radio and X-ray data can be used to characterise the relativistic jets \citep[e.g.][]{An2020,Spingola2020,Ighina2022a}.

Even though the number of $z>6$ RL QSOs has been limited to a handful of sources for several years (e.g., \citealt{McGreer2006,Willott2010}), recent efforts significantly increased the number of RL sources at high redshifts (e.g., \citealt{Belladitta2020,Banados2021}), including obscured radio QSOs \citep[e.g.,][]{Drouart2020,Endsley2022,Endsley2023}, potentially up to $z\sim7.7$ \citep{Lambrides2024}.
The discovery of many $z>6$ RL QSOs has also been possible thanks to the recent advent of new-generation radio surveys. Two examples of such surveys are the LOFAR Two-metre Sky Survey \citep[LoTSS;][]{Shimwell2017,Shimwell2019,Shimwell2022} and the Rapid ASKAP Continuum Survey \citep[RACS:][]{McConnell2020,Hale2021,Duchesne2023}. 
Both these surveys enabled the radio detection of already known high-$z$ QSOs as well as the discovery of new ones (see, e.g., \citealt{Gloudemans2021,Gloudemans2022} for LOFAR and \citealt{Ighina2021a,Ighina2023} for RACS).

In particular, in \cite{Ighina2021a}, we were able to uncover the RL nature of the $z=6.44$ QSO VIKING J231818.3$-$311346 (hereafter VIK~J2318$-$31) thanks to the first data release of the RACS-low survey, at 888~MHz \citep{McConnell2020}. Being one of the most distant RL QSOs currently known, 
the characterisation of VIK~J2318$-$31 and the other few similar sources in this redshift range can provide more stringent constraints on the evolution of SMBHs hosted in RL systems than the majority of the other sources in this class. Moreover, in \cite{Ighina2022b}, by combining radio observations of the Galaxy And Mass Assembly (GAMA; \citealt{Driver2011}) 23$^{\rm h}$ field, where VIK~J2318$-$31 is located, we constrained its radio spectrum over two orders of magnitudes in frequency ($\sim$0.1--10~GHz, observed, corresponding to $\sim$0.7--75~GHz in the rest frame). Based on the shape of the radio spectrum and on its variability at 888~MHz, we found that this system is likely hosting a young radio jet with a projected size of the order of milli-arcsec (mas; $\sim$5--10~pc). Similar conclusions were also drawn from very long baseline interferometry (VLBI) observations \citep{Zhang2022}, where the source is only partially resolved on scales of $\sim2$~mas. 

In this work, we present a multi-wavelength study of this specific source covering the radio, optical/IR and X-ray bands. The work is based on newly obtained observations as well as archival ones not yet published. In Sec. \ref{sec:data} we describe the data underlying  this work. In Sec. \ref{sec:disc} we use radio and X-ray observations in order to constrain the properties of the relativistic jets and optical/near infrared (NIR) observations to estimate the mass of the central BH. Finally, we summarise our result in Sec. \ref{sec:conclusions}.

Throughout the paper we assume a flat Lambda cold dark matter ($\Lambda$CDM) cosmology with $H_{0} = 70$\,km\,s$^{-1}$\,Mpc$^{-1}$, $\Omega_\mathrm{M}=0.3$, and $\Omega_{\Lambda}=0.7$, where 1\arcsec\, corresponds to a projected distance of 5.49~kpc at $z=6.44$. Spectral indices are given assuming $S_{\nu}\propto \nu^{-\alpha}$ and all errors are reported at 68\% confidence unless specified otherwise.

\section{Multi-wavelength data}
\label{sec:data}

VIK~J2318$-$31 has already been targeted with several radio observations, described in \cite{Ighina2022b} and \cite{Zhang2022}, as well ALMA observations, see \cite{Decarli2018}, \cite{Venemans2020} and \cite{Neeleman2021}. In this Section, we describe datasets that have not been published yet in the X-ray, radio and optical/NIR bands.

\subsection{X-ray: Chandra}

\begin{figure*}
\centering
	\includegraphics[width=0.45\hsize]{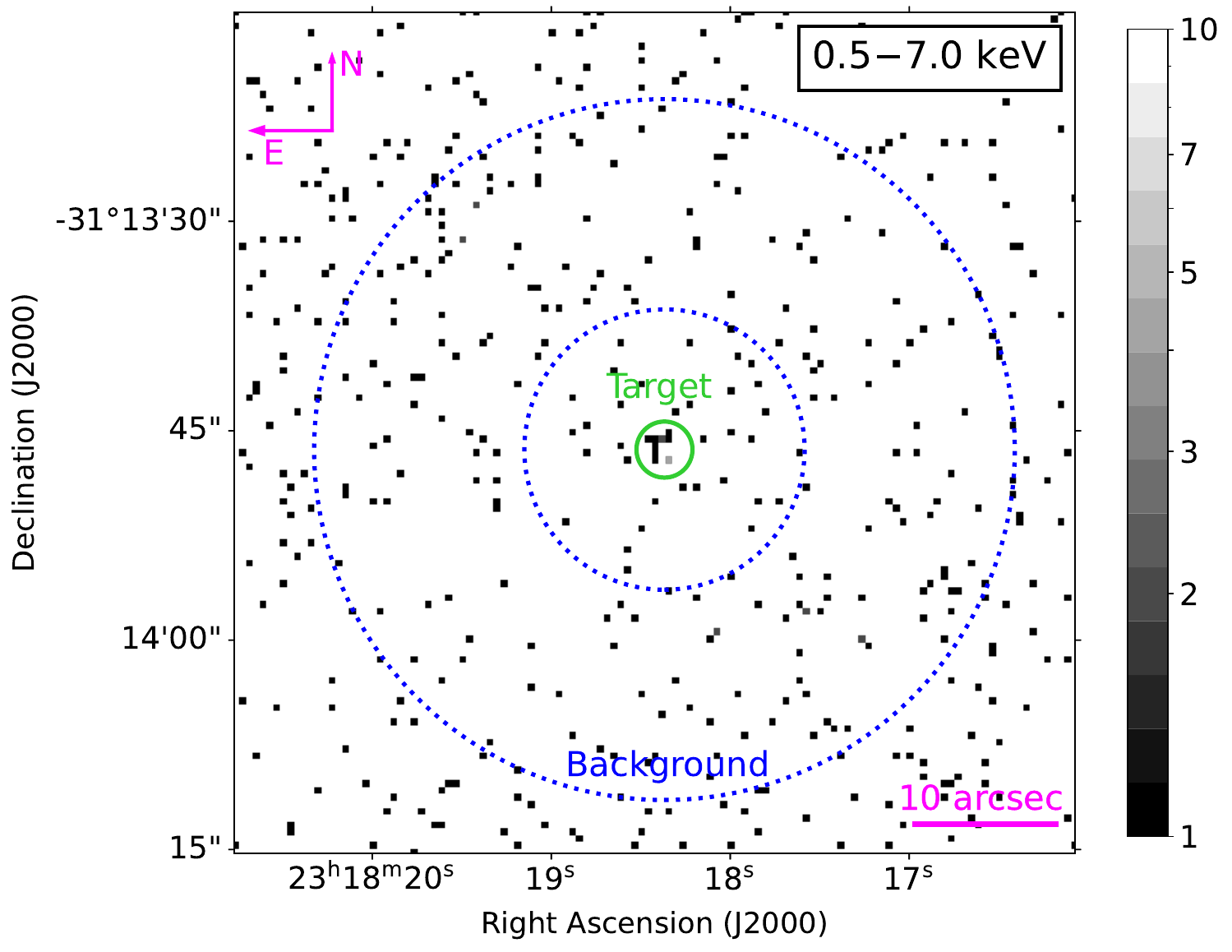}
	\includegraphics[width=0.49\hsize]{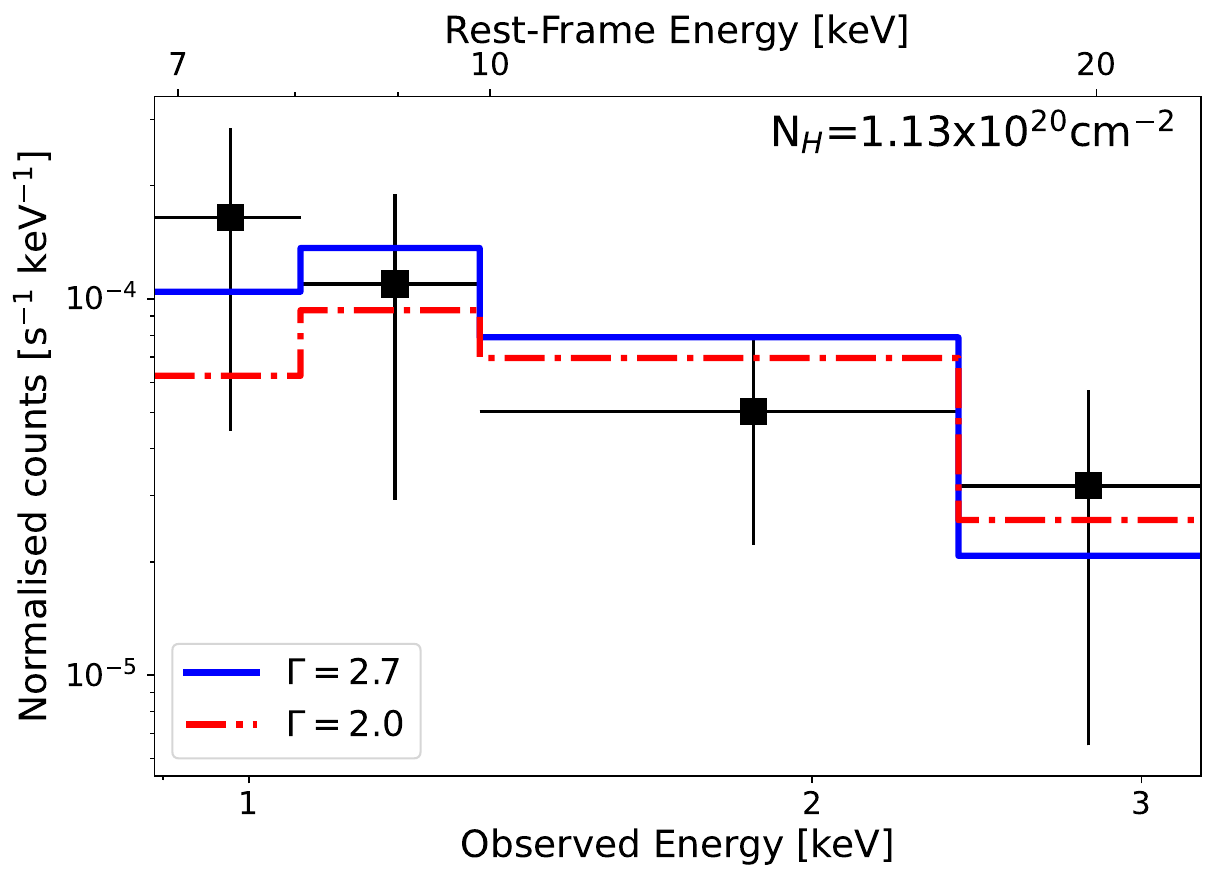}
    \caption{{\bf Left:} X-ray image of VIK~J2318$-$31 from a 70~ksec Chandra observation. The extraction region for the target and background spectra are reported as a green solid circle and a blue dotted annulus. {\bf Right:} X-ray spectrum of VIK~J2318$-$31. The best-fit power law spectra obtained using the \textit{cstat} statistics are reported as a solid blue line (with $\Gamma$ as a free parameter) and as a dashed-dotted red line (with $\Gamma$ fixed). The corresponding best-fit values are reported in Tab. \ref{tab:x-ray_values}.}
    \label{fig:x-ray_im_spec}
\end{figure*}

VIK~J2318$-$31 was observed with the Chandra X-ray telescope during Cycle 23 for a total of 70~ksec (project 704300, P.I. L. Ighina). Data reduction was performed using the Chandra Interactive Analysis of Observations (\texttt{CIAO}; \citealt{Fruscinone2006}) software package (v4.13) with CALDB  (v4.9.5).
We used the \texttt{SPECEXTRACT} task in order to extract the photons from a 2\arcsec \, region centred on the optical/NIR position of the QSO, while the background photon counts from a 10$-$25\arcsec \, annulus, again centred at the optical/NIR position of the QSO (see Fig. \ref{fig:x-ray_im_spec}). The source is detected with 12 net photons over an expected background of $\sim$1 photon.
We then analysed the extracted photons using the \texttt{XSPEC} (v12.11.1) package and performed a fit adopting the C-statistic \citep{Cash1979} in the energy range 0.5--7~keV (in order to reduce the background noise) with a power law absorbed by the Galactic column density along the line of sight (N$_{\rm{H}}=1.13\times10^{20}$~cm$^{-2}$; \citealt{HI4PI2016}). In particular, we considered both a model where the photon index value was fixed to $\Gamma=2.0$ (as typically observed in $0<z<6$ QSOs \citep[see, e.g.,][]{Vignali2005, Piconcelli2005, Nanni2017} and one where it was free to vary. We report in Tab. \ref{tab:x-ray_values} and Fig. \ref{fig:x-ray_im_spec} the results of the analysis, together with the $\tilde{\alpha}_{\rm ox}$ parameter value, following \cite{Ighina2019}. This parameter is defined as $\tilde{\alpha}_{\rm ox}$=--0.303 log$\frac{L_{10keV}}{L_{2500\text{\normalfont\AA}}}$ and is meant to quantify the relative emission in the X-rays, produced by the jets and/or the X-ray corona, compared to the optical one, associated with the accretion disc. $\tilde{\alpha}_{\rm ox}$ is related to the commonly used ${\alpha}_{\rm ox}$, defined at 2~keV instead of 10~keV, as $\tilde{\alpha}_{\rm ox}$= 0.79 $\times \, {\alpha}_{\rm ox}$ + 0.21 $\times$ ($\Gamma$ -- 1.0).

\begin{table}
	\centering
	\caption{Best-fit values obtained from the fit of the X-ray spectrum of VIK~J2318$-$31 assuming a simple power law with Galactic absorption (N$_{\rm H}=1.13\times10^{20}$~cm$^{-2}$; \citealt{HI4PI2016}). Col. (1) best-fit photon index (top) and fixed value (bottom); col. (2) un-absorbed flux in the energy band 0.5--10~keV; col. (3) rest-frame luminosity in the energy range 2--10~keV; col. (4) $\tilde{\alpha}_{\rm ox}$ parameter; col. (5) c-statistic and degrees of freedom of the fit.}
	\label{tab:x-ray_values}
\begin{threeparttable}
\centering
\begin{tabular}{ccccccc}
\hline
\hline
$\Gamma$    &   f$_\mathrm{0.5-7keV}^\mathrm{a}$ &   L$_\mathrm{2-10keV}^\mathrm{b}$   &   $\tilde{\alpha}_{\rm ox}$  & cstat/d.o.f.\\
\hline
2.7$^{+1.1}_{-1.0}$  & 	3.3$_{-0.9}^{+1.8}$  &   1.7$_{-1.3}^{+4.1}$  & 1.42$_{-0.04}^{+0.05}$ & 8.7 / 10 \\
\hline
2.0$^c$ & 3.2$^{+0.9}_{-0.9}$ & 0.9$^{+0.6}_{-0.4}$ & 1.44$_{-0.04}^{+0.04}$ & 9.2 / 11 \\
\hline
\hline
\end{tabular}

\begin{tablenotes}
\item Errors are reported at 1{\rm $\sigma$} level of confidence.
\item [a] in units of 10$^{-15}$ erg s$^{-1}$ cm$^{2}$
\item [b] in units of 10$^{45}$ erg s$^{-1}$
\item [c] assumed to vary by $\pm0.5$ in the flux and luminosity uncertainty computation
\end{tablenotes}

\end{threeparttable}
\end{table}

\subsection{Radio observations}
In \cite{Ighina2022b} we already constrained the radio emission of VIK~J2318$-$31 using multiwavelength data obtained as part of surveys or dedicated observations. However, we note that, after a more detailed analysis of the 888~MHz ASKAP image of the G23 field taken in March 2019, \cite{Gurkan2022} corrected the flux density of VIK~J2318$-$31 to be S$_{\rm 888MHz}=0.68\pm0.07$~mJy~beam$^{-1}$ with respect to the one we reported in \cite{Ighina2021a,Ighina2022b}. In the following, we adopt the updated value from \cite{Gurkan2022}. 

Moreover, after the publication of \cite{Ighina2022b}, more observations of the G23 field have been performed at 216~MHz with the Murchinson Widefield Array (MWA; \citealt{Tingay2013}) as part of the  MWA Interestingly Deep AStrophysical (MIDAS) Survey (Paterson et al., in preparation). This new 216~MHz image uses 1701 2-min snap-shot observations (compared to the 154 used in the pilot MIDAS image previously described in \cite{Quici2021}. 
Additionally, an improved pipeline has been used for the new analysis which calibrates from the GaLactic and Extra-Galactic All-sky MWA (GLEAM; \citealt{Wayth2015, Hurley2017}) survey and also processes the two polarisations separately. Numerous improvements are incorporated for this ultra-deep MWA field with full details available in Paterson et al. (in preparation). We report in Fig. \ref{fig:MWA} the cutout of these observations around VIK~J2318$-$31 where an RMS of 0.28 mJy~beam$^{-1}$ is reached (with respect to 0.45~mJy~beam$^{-1}$ reported in \citealt{Ighina2022b}). 
As this new MWA image is close to being confusion limited, this new RMS is measured by fitting a Gaussian to the peak of the pixel distribution. Interestingly, VIK~J2318$-$31 is still not detected even in this new, deeper image (although the peak flux is 0.51~mJy~beam$^{-1}$ = 1.8$\times \sigma$).
The flux scale is accurate to $\sim5$\% with respect to GLEAM from cross-matching of bright and isolated sources. We include this value, added in quadrature, when computing the 3$\sigma$ upper limit for the low-frequency radio emission of VIK~J2318$-$31 (see Tab. \ref{tab:VLA_fit}).

\begin{table*}
	\centering
	\caption{Best-fit flux densities obtained from a Gaussian fit to the radio images of VIK~J2318$-$31 together with the synthesised beam size, Position Angle (P.A.) and RMS of each observation. The upper limits reported in this table are at 3$\sigma$}
\label{tab:VLA_fit}
\begin{tabular}{cccccccc}
\hline
\hline
Epoch & Telescope & Frequency & Int. flux & Peak surf. & Beam size & P.A. East & Off-source \\
 & & & density & brightness & & of North & RMS\\
& & (GHz) & (\textmu Jy) & (\textmu Jy~beam$^{-1}$) &  (maj\arcsec$\times$min\arcsec) & (deg) & (\textmu Jy~beam$^{-1}$)\\
\hline
March--August\\
2018 & MWA & 0.216 & $<860$ & $<860$ & 63.8$\times$53.7 &$-$37.0 & 280\\
\hline
April & uGMRT & 0.400 & 990$\pm$206 & 985$\pm$123 & 9.4$\times$5.7 &$-$1.7 & 113\\
2022 & uGMRT & 0.650 & 787$\pm$60 & 782$\pm$46 & 7.1$\times$3.8 & 30.0 & 26\\
& ATCA  & 2.1 & 565$\pm$57 & 504$\pm$35 & 13.8$\times$5.6 & $-$17.1 & 24\\
& ATCA  & 5.5 & 155$\pm$23 & 149$\pm$14 & 8.1$\times$3.6  & $-$0.4 & 12\\
& ATCA  & 9   & --- & 54$\pm$10 & 4.2$\times$2.3 & 0.0 & 9\\
\hline
September & uGMRT & 0.400 & 1070$\pm$260 & 791$\pm$132 & 8.4$\times$5.0 & $-$26.4 & 123 \\
2022 & uGMRT & 0.650 & 798$\pm$59 & 786$\pm$48 & 8.5$\times$3.1 & 11.0 & 28 \\
& ATCA  & 2.1 & 409$\pm$41 & 398$\pm$28 & 10.6$\times$4.1 & 2.3 & 20\\
& ATCA  & 5.5 & 138$\pm$34 & 143$\pm$20 & 4.8$\times$1.9  & 4.4 & 19\\
& ATCA  & 9   & $<79$ & $<79$ & 3.0$\times$1.1 & 4.0 & 25\\


\hline
\hline
\end{tabular}
\end{table*}

Dedicated upgraded Giant Metrewave Radio Telescope (uGMRT) observations of VIK~J2318$-$31 (project 42\_001; P.I. L. Ighina) were performed in two different seven hour epochs: 2022 April 22 and 2022 September 02. In both epochs the target was observed in band-3 (centred at 400~MHz) and in band-4 (centred at 650~MHz) using the GMRT wideband backend (GWB; \citealt{Reddy2017}). During each run, we observed sources 3C48 and 0025$-$260 as primary and secondary calibrators, respectively. For the data reduction of these observations we used the CAsa Pipeline-cum-Toolkit for Upgraded GMRT data REduction (\texttt{CAPTURE}; \citealt{Kale2021}) code, which is based on the Common Astronomy Software Applications package \citep[\texttt{CASA};][]{Mcmullin2007}, and applied further flagging when needed. In particular, during the imaging part, which relies on the \texttt{tCLEAN} task, we adopted a robust parameter of 0.5. We report the images obtained in Fig. \ref{fig:uGMRT} and the results of the 2D Gaussian fit (with \texttt{CASA}) in Tab.~\ref{tab:VLA_fit}. We consider a further 5\% in the uncertainties (added in quadrature) of the measurements derived from these images in order to account for the uncertainty related to the calibration process.

Similarly to uGMRT, Australia Telescope Compact Array (ATCA) observations were performed in two epochs (project C3477; P.I. L. Ighina) of 12 hours each and simultaneous to the uGMRT ones: 2022 April 22 (configuration 1.5A) and 2022 September 02 (configuration 6D). Observations were carried at 2.1, 5.5 and 9~GHz using the Compact Array Broadband Backend (CABB; \citealt{Wilson2011}), with a nominal bandwidth of 2048~MHz (divided in channels of 1~MHz). We used the source PKS~B1934$-$638 as the standard primary calibrator \citep{Reynolds1994}, observed at the beginning of each session, and J2255$-$282 as secondary calibrator, alternating its observations with the target in order to ensure the phase calibration throughout the runs. To process the data (calibration and imaging), we used the \texttt{MIRIAD} data-reduction package \citep{Sault1995} following a standard reduction. In particular, for imaging, we adopted a robust parameter of 0.5. 
Finally, we performed a 2D Gaussian fit on the target using \texttt{CASA}. The final images are reported in Fig. \ref{fig:ATCA}. VIK~J2318$-$31 is detected in all the observations apart from the September run at 9~GHz. Moreover, the singal-to-noise (S/N) in the 9~GHz image from April is not enough to perform a fit to the source. Therefore, in this last case we only report the peak emission. We show the results of the fits in Tab. \ref{tab:VLA_fit}, where we also considered a conservative 5\% (added in quadrature) in the errors of these ATCA measurements in order to account for the uncertainty related to the calibration process. We note that the different sensitivities reached during the two observing runs are mainly due to poor weather conditions during the September run, which affected the higher frequencies more. In particular, during the September run the median seeing was a factor $\sim$1.5 larger than in the April run (240~\textmu m and 160~\textmu m, respectively).  At the same time, the more extended configuration in the September run significantly reduced radio frequency interference (RFI) at 2.1~GHz, thus increasing the sensitivity at this frequency compared to the previous run. In all the images, the source is unresolved.

\subsection{Optical/NIR: X-Shooter and GNIRS}

\begin{figure*}
\centering
	\includegraphics[width=0.9\hsize]{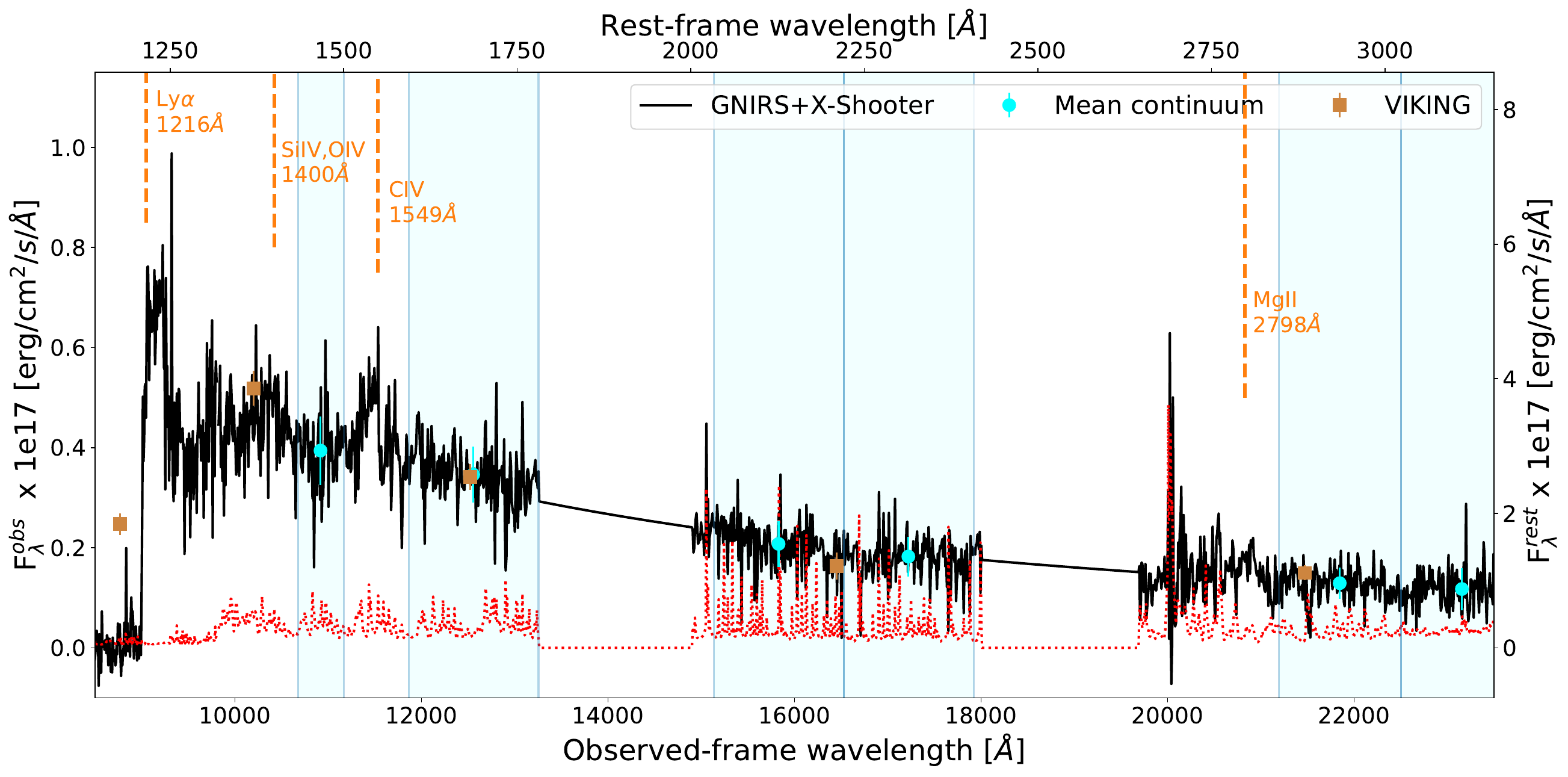}
    \caption{NIR spectrum of VIK~J2318$-$31 obtained from GNIRS+X-Shooter observations (black continuum line). The spectrum was re-binned to a resolution of 250~km~sec$^{-1}$. The dotted red line indicates the error on the spectrum. The regions affected by strong telluric absorption have been masked. The cyan shaded areas indicate the wavelength regions used to derive photometric points (cyan circles) used for the accretion model fit. The dashed orange lines indicate the expected wavelength of the brightest emission lines present in the observed spectrum.}
    \label{fig:VIK_NIR_spec}
\end{figure*}

Archival optical/NIR spectroscopic observation with X-Shooter (Very Large Telescope; \citealt{Vernet2011}) and with the Gemini near-infrared spectrograph (GNIRS; Gemini-North Telescope; \citealt{Elias2006b,Elias2006a}) of VIK~J2318$-$31 are available. The target was observed with GNIRS for a total of 1~h (12 segments of 300s each) with a 1\arcsec slit (program id: GN-2013B-Q-50, P.I. K. Chambers) on the 2013 September 30. Moreover, the source was observed for a total of 2.7~h with X-Shooter (8 segments of 1200s each) with a 0.9/0.6\arcsec slits for the OPT/NIR arm (program id: 097.B-1070, P.I. Farina), half performed on 2016 August 05 and half on the 2016 August 25. For the data reduction of the X-shooter VIS data, we used the \texttt{ESOReflex} workflow \citep{Freudling2013}.
For the data reduction of the GNIRS and X-shooter-NIR observations we used the Python Spectroscopic Data Reduction Pipeline (\texttt{PypeIt}) Python package  (\citealt{Prochaska2020a,Prochaska2020b}; see, e.g., \citealt{Banados2021,Banados2023} for applications on similar objects). In particular, we only consider half of the X-shooter-NIR segments (from the 2016 August 05 run) since during the other half the seeing was significantly worse (1.2\arcsec) with respect to the first run (0.6\arcsec) and the loss in flux, for an already faint target, was too severe for an optimal spectrum extraction. We extracted 1D spectra from each observing segment, we combined the spectra from the same instrument and then fitted a telluric model based on the grids from the Line-By-Line Radiative Transfer Model LBLRTM4, \citep{Clough2005, Gullikson2014}. In order to have a consistent absolute flux calibration, we normalised both the GNIRS and X-shooter-NIR spectrum to the VIKING-band-J magnitude, while the X-shooter-VIS spectrum was normalised to the VIKING-band-Z magnitude. Both magnitudes have been corrected for Galactic extinction assuming $R_{\rm V}=3.1$ \citep{Schlafly2011} and the extinction law from \cite{Fitzpatrick1999}. In both cases the corrections were $\Delta{\rm mag}<0.02$. We then re-binned the spectra to a common wavelength grid with a pixel size of 250~km~sec$^{-1}$ (using the SpectRes: Simple Spectral Resampling, \texttt{SpecRes}, code; \citealt{Carnall2017}) and we averaged them using the inverse variance of each spectrum as weight. We report the spectrum obtained in Fig. \ref{fig:VIK_NIR_spec}, where we removed the wavelengths affected by heavy telluric absorption. In the following we adopt the redshift derived from the [CII] emission line (158 \textmu m, rest frame) observed in ALMA as the most reliable estimate: 6.4429$\pm$0.0003 \citep{Venemans2020}.

\section{Multi-wavelength properties}
\label{sec:disc}
VIK~J2318$-$31 is one of the very few RL QSOs at $z>6$ for which observations along the entire electromagnetic spectrum are available. In the following, we discuss its properties based on the multi-wavelength data described in the previous section. 

\subsection{X-ray emission}

\begin{figure}
	\includegraphics[width=\hsize]{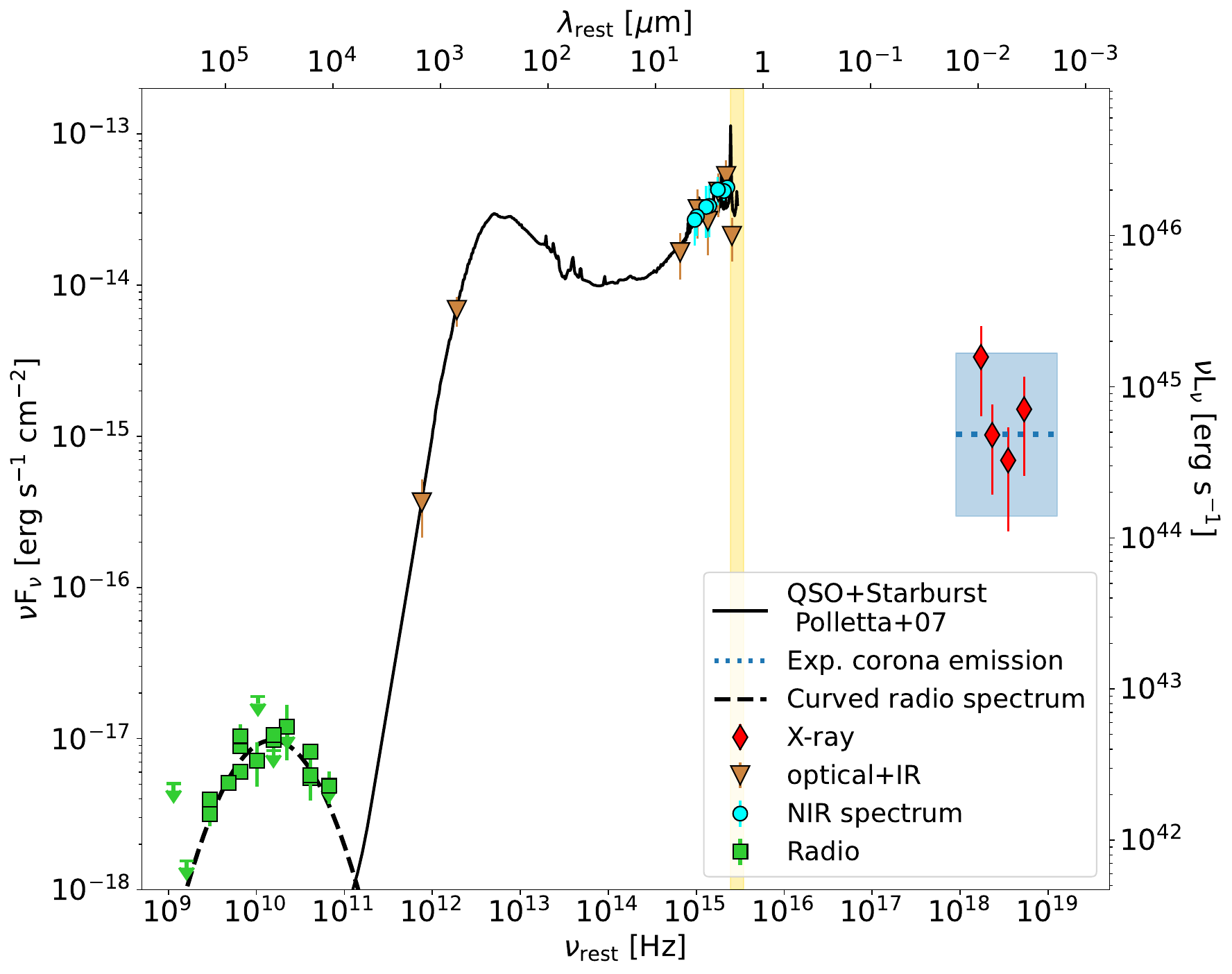}
    \caption{Rest-frame spectral energy distribution (SED), from the radio to the X-ray band, of VIK~J2318$-$31. Radio data are reported as green squares together with the best-fit curved radio spectrum derived (dashed black line); optical and IR data from the literature are reported as brown inverted triangles, while the photometric points used for the accretion disc model are reported as cyan circles; the X-ray Chandra data are reported as red diamonds together with the expected X-ray emission from the L$_{\rm X}-{\rm L}_{2500\text{\normalfont\AA}}$ relation derived in {\protect\cite{Lusso2016}} (assuming $\Gamma=2$; blue shaded region). The shaded yellow area represents the range of optical frequencies where the Ly$\alpha$ absorption from the intergalactic medium (IGM) takes place.}
    \label{fig:SED_VIK}
\end{figure}

The X-ray emission in RL QSOs is a useful tool to assess the orientation of their relativistic jets, or in other words, their blazar nature (e.g., \citealt{Ghisellini2015c}). Indeed, if one of the relativistic jets is oriented close to our line of sight (i.e., in the blazar case), we expect relativistic beaming effects to significantly increase the X-ray emission (even more than the radio one, e.g., \citealt{Ghisellini2013b}) as well as to produce a `flat' X-ray spectrum ($\Gamma \lesssim 1.7$; e.g., \citealt{Ighina2019,Paliya2020b}). Based on the best-fit parameters derived for VIK~J2318$-$31 (see Tab. \ref{tab:x-ray_values}), its relativistic jets are unlikely to be oriented close to our line of sight. Indeed, a best-fit photon index value of $\Gamma = 2.7^{+1.1}_{-1.0}$ and an X-ray-to-optical ratio $\tilde{\alpha}_{\rm ox} = 1.42^{+0.05}_{-0.04}$ (or  $\tilde{\alpha}_{\rm ox} = 1.44\pm{0.04}$, assuming $\Gamma=2.0\pm0.5$) are significantly larger than what is normally observed in blazars, even at high redshifts ($\Gamma \lesssim 1.7$ and $\tilde{\alpha}_{\rm ox} \lesssim 1.34$; see, e.g., \citealt{Ighina2019}). Moreover, the overall X-ray luminosity is fully consistent with the one expected from the $\rm L_{\rm 2500\text{\normalfont\AA}}-L_{\rm X}$ relation derived by \cite{Lusso2016} (assuming $\Gamma=2$; see Fig. \ref{fig:SED_VIK}) and then also confirmed at high redshift \citep[e.g.][]{Vito2019,Salvestrini2019}. It is therefore likely that the X-ray emission observed in VIK~J2318$-$31 is dominated by the X-ray corona rather than a relativistic jet. 

At the same time, by considering the bolometric luminosity of VIK~J2318$-$31 derived from the optical emission ($L_{\rm bol}=8.4\pm2.8\times10^{46}$~erg~sec$^{-1}$ see subsection \ref{sec:SMBH}), we can compute the X-ray bolometric correction associated with the VIK~J2318$-$31 system, defined as the ratio $K_{\rm X}=L_{\rm bol}/L_{\rm 2-10keV}$. In this way, we obtain $K_{\rm X}\sim50-90$ (depending on the value of $\Gamma$ assumed), which is consistent with the value derived, for example, by \cite{Shen2020} ($K_{\rm X}=110^{+70}_{-40}$). Once again, this suggests that the contribution from the relativistic jet to the observed X-ray emission in VIK~J2318$-$31 is only marginal. 

\subsection{Radio properties}
\begin{figure}
	\includegraphics[width=\hsize]{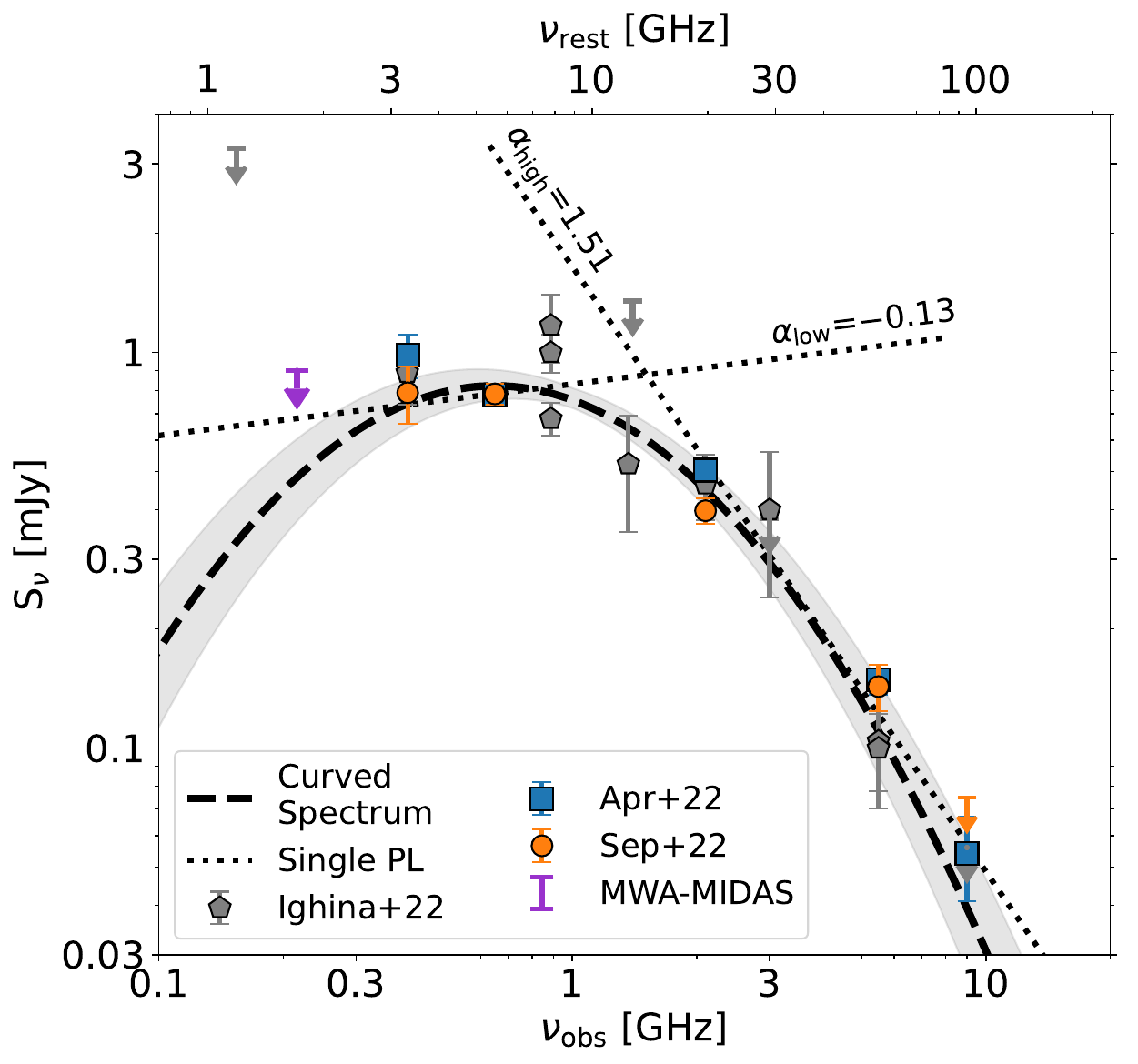}
    \caption{Radio spectrum of VIK~J2318$-$31. Data already presented in {\protect\cite{Ighina2022b}} are shown as grey pentagons, while new observations are reported as blue squares (April epoch) and orange circles (September epoch). The purple upper-limit is the estimate derived from the new MWA-MIDAS image. The dashed black line shows the best-fit curved spectrum described in the text. The shaded grey area encloses the curves obtained by varying the best-fit parameters by $\pm1\sigma$. The dashed line shows the best-fit double power law, while the dotted lines are the best-fit single power law describing the data above and below 1.4~GHz in the observed frame.}
    \label{fig:radio_spec}
\end{figure}

In our previous studies focused on the radio properties of VIK~J2318$-$31, we found that the radio spectrum shows a flattening at lower frequencies and signs of variability at 888~MHz ($\sim$2.5$\sigma$ previously reported and reduced to $\sim$2$\sigma$ with the new analysis of the GAMA 23 field; \citealt{Gurkan2022}). In Fig. \ref{fig:radio_spec} we report all the radio measurements available for VIK~J2318$-$13 from \cite{Ighina2022b} together with the new data discussed in Sec. \ref{sec:data}. From the newly obtained simultaneous observations with ATCA and uGMRT, we do not find any further variability in the observed range of 0.4--9~GHz: all the measurements are consistent within $\sim$2$\sigma$. 
At the same time, these uGMRT+ATCA simultaneous radio data ultimately confirm the presence of a flattening of the radio spectrum at low frequencies. Therefore, similarly to \cite{Ighina2022b}, we performed a fit, in the rest frame, to all the radio data points (Fig. \ref{fig:radio_spec}) using the \texttt{MrMOOSE} code \citep{Drouart2018b,Drouart2018a} with a curved radio spectrum of the form:
\begin{equation}
    S_\nu = N \, \nu^{-\alpha} \, e^{q\,({\mathrm{ln}} \nu)^2},
    \label{eq:curved_pl}
\end{equation}

where $N$ is the normalisation, $\alpha$ is the index of the power law and the $q$ parameter is a measurement of the curvature of the spectrum ($|q|>0.2$ indicates a significantly curved spectrum; \citealt{Callingham2017}). The peak frequency of the spectrum is given by $\nu_\mathrm{peak}$ = $e^{-\alpha/2q}$. 
In this way we obtained $q=0.44\pm0.4$ and $\alpha=0.38\pm0.06$, which result in a peak frequency of $\nu_{\rm peak}= 650\pm60$~MHz, consistent with the values previously found. 
Even though in \cite{Ighina2022b} we also discussed the possibility of a double power law as a potential radio spectral model due to the cooling of the most energetic electrons with the jets on timescales of $10^{4-5}$~yr, recent Very Large Baseline Array (VLBA) observations of VIK~J2318$-$31 revealed that the radio emission is very compact ($\sim$2~mas) and, therefore, likely associated with a very young radio jet \citep[see][]{Zhang2022}, with a kinematic age of $\sim$200~yr (assuming an expansion velocity of $\sim$0.2~$c$; e.g., \citealt{An2012}).

Nevertheless, in order to ease the comparison of our source with other high-$z$ sources from the literature with a more limited frequency coverage, we also performed a fit to the radio spectrum of VIK~J2318$-$31 considering two single power laws, describing the emission at frequencies above and below $\sim1.4$~GHz in the observed frame. 
In this way we obtained a spectrum with $\alpha_{\rm low}=-0.13\pm0.09$ for $\nu\lesssim1$~GHz and with $\alpha_{\rm high}=1.51\pm0.11$ at larger frequencies, see Fig. \ref{fig:radio_spec}. These values are similar to what is normally found in other $z>5$ radio objects, both in the $0.14-1.4$~GHz frequency range (e.g., \citealt{Gloudemans2021}) as well as at $\sim1.4-10$~GHz (e.g., \citealt{Drouart2020,Shao2022}).
Also in this case, the different signs of the spectral indexes at low and high frequencies indicate the presence of a turnover in the radio spectrum of VIK~J2318$-$31 around $\sim1$~GHz in the observed frame. 
As described below, such a peak can be understood as due to synchrotron self-absorption (SSA) and/or free-free absorption (FFA).

Having an estimate of the turnover frequency as well as the linear size of the radio source (from \citealt{Zhang2022}; $\sim$2.3~mas or $\sim$13~pc at $z=6.44$) we can compare VIK~J2318$-$31 to the empirical linear size-peak frequency relation found in the peaked spectrum (PS) radio sources \citep[e.g.,][]{Odea2021}. Using the peak-size relation found by \cite{Orienti2014}, which was derived using objects that cover a wide range of linear sizes ($\sim0.01-100$~kpc, see their fig. 6), a source 13~pc in size would have the peak of its radio emission around 8~GHz in the rest frame. 
If we also consider the fact that the jet of VIK~J2318$-$31 is probably seen at an angle $\sim$30$^\circ$, based on the fact that the this source is an un-obscured QSO (i.e., we expect $\theta \lesssim45^{\rm o}$; e.g.,  \citealt{Urry1995}) and that the X-ray and radio observations are not compatible with a blazar nature (i.e., $\theta \gtrsim10^{\rm o}$; e.g., \citealt{Ghisellini2015b}), the expected peak frequency would be 5.3~GHz.
These values are in good agreement with the estimate obtained from the fit of the radio spectrum, $\sim$5~GHz. Since the peak frequency -- linear size relation can be well reproduced by the SSA model \citep{Kellermann1981}, the good agreement between our data and the \cite{Orienti2014} relation could suggest that the decrease in flux below $\sim$650~MHz is caused by SSA. 

Another way to test the origin of the peak in the radio spectrum of PS sources is to compare the magnetic field estimated from the peak frequency, assuming it is produced by SSA, and the one derived assuming equipartition (between radiative particles and magnetic field). A significant discrepancy between the two values indicates that the turnover in the radio spectrum is not originated by SSA \citep[e.g.,][]{Keim2019}. 
Under the assumption of SSA, the values of the angular size ($\theta_{\rm src, maj}$ and $\theta_{\rm src, min}$), the peak frequency ($\nu_{\rm peak}$) and the flux density at the peak frequency (S$_{\rm peak}$) can be used to infer the magnetic field strength using the following equation:
\begin{equation}
    B_{\rm SSA} \approx \frac{[\nu_{\rm peak}/ f(\alpha_{\rm thin)}]^5 \, \theta_{\rm src, maj}^2\, \theta_{\rm src, min}^2}{S_{\rm peak}^2 \, (1+z)},
    \label{eq:Bssa}
\end{equation}

where the magnetic field is in Gauss, $\nu_{\rm peak}$ in GHz, S$_{\rm peak}$ in Jy and $\theta_{\rm src, maj}$ and $\theta_{\rm src, min}$ are in mas. All the values are in the observed frame. The function $f(\alpha_{\rm thin})$ is as defined by \cite{Kellermann1981} and we fix it to $f(\alpha_{\rm thin}) = 8$, as typically assumed in the literature \citep[e.g.,][]{Orienti2008,Ross2023}.

The value of the magnetic field assuming equipartition mainly depends on the intensity (i.e., luminosity) and volume of the emitting region \citep{Pacholczyk1970}. In particular, following eq. A.6 in \cite{Spingola2020}, the equipartition magnetic field is given by:
\begin{equation}
        B_{\rm eq} \approx  \left(4.5 \, (1+\eta) \, c_{12} \frac{L}{V} \right)^{2/7}, 
        \label{eq:Beq}
\end{equation}

where we assumed $\eta = 1$, that is, electrons and protons contribute equally to the overall energy, $c_{12} = 3.9 \times 10^7$ is a constant, L is the luminosity at the rest-frame frequency of 20~GHz ($\sim$2.5~GHz observed; i.e. in the optically thin part of the spectrum) and V is the volume of the radio-emitting region in cm$^3$. For this last parameter we assume an ellipsoidal geometry with the major axis given by $\theta_{\rm src, maj}$ and the two minor axis equal to $\theta_{\rm src, min}$.

Even though we have a good estimate of the peak frequency and the emission across a wide range of frequencies, the size of the radio emitting region in VIK~J2318$-$31 is poorly constrained, having only one measurement along the major axis from the VLBA observations \citep{Zhang2022}. However, since there is only one unconstrained parameter, $\theta_{\rm src, min}$, for two equations, we can compute the dimensions of the jets along their minor axis to give a similar magnetic field value. In particular, in order obtain the same magnetic field from eq. \ref{eq:Bssa} and eq. \ref{eq:Beq} ($B\sim0.2$~mG) a jet with a minor axis of $\sim 7$~\textmu as is required. In order to have values consistent within a factor $\sim3$, the minor axis must be $\lesssim 10$~\textmu as. These values are orders of magnitude smaller than the size of a relativistic jets in PS objects (a few mas, e.g., \citealt{Dallacasa2000,Orienti2008,Orienti2014,Keim2019,Ross2023}). It is therefore unlikely that the turnover present in the radio spectrum of VIK~J2318$-$31 can be explained by SSA.
To ultimately rule out SSA as the cause of the low-frequency absorption in VIK~J2318$-$31 a good sampling of the optically thick part is needed in order to model the overall spectrum. Indeed, from the accurate model fitting of ten $z>5$ RL QSOs, \cite{Shao2022} found that the peak in the radio spectrum of the majority of their sources can be better described by a FFA model with respect to a SSA one. 

This result may suggest that FFA absorption is more frequent in high-$z$ systems compared to the more local Universe.

\subsection{Optical/NIR data: SMBH mass and accretion}
\label{sec:SMBH}

In this section we use optical and NIR data in order to estimate the mass of the BH at the centre of the VIK~J2318$-$31 system. In particular, we used two methods: a single epoch (SE) method, based on the width of broad emission lines (CIV1549\AA \, and MgII2798\AA), and a fit of the rest-frame UV continuum emission with an accretion disc model.

\subsubsection{Single Epoch estimate}
One of the most common methods to estimate the mass of the SMBH hosted by high-$z$ QSOs is to use the the broad emission lines observed in their optical-NIR spectrum. In particular, under the assumption that the gas present in the Broad Line Region (BLR) is in virial motion around the SMBH and that the distance of the BLR from the central BH depends on the disc luminosity ($\propto$L$_{\rm disc}^{\alpha}$, with $\alpha\sim0.5$; e.g., \citealt{Kaspi2000,Bentz2009}), the width of the broad emission lines and the continuum optical luminosity can be used as proxies for the mass of the central object \citep[e.g.,][]{Vestergaard2006,Vestergaard2009}. In the following we focus on two specific broad emission lines accessible at high redshift with ground-based spectrographs, CIV1549 and MgII2798 \cite[e.g.,][]{Shen2019,Belladitta2022,Mazzucchelli2023}.

To model the continuum emission in the NIR spectrum we followed the prescription detailed in \cite{Mazzucchelli2017}. We considered three components: a power law, a Balmer-pseudo-continuum and an iron pseudo-continuum. Similarly to other studies of high-$z$ QSOs, we fixed the emission from the Balmer pseudo-continuum (see, e.g. eq. 2 in \citealt{Banados2021}) to be 30\% that of the power law at rest-frame 3646\AA. Moreover, we considered the iron template derived by \cite{Vestergaard2001} and convolved it with a Gaussian function with the same width as the MgII line, assuming that the FeII and the MgII emission lines are produced by nearby regions.

\begin{figure*}
\centering
	\includegraphics[width=0.434\hsize]{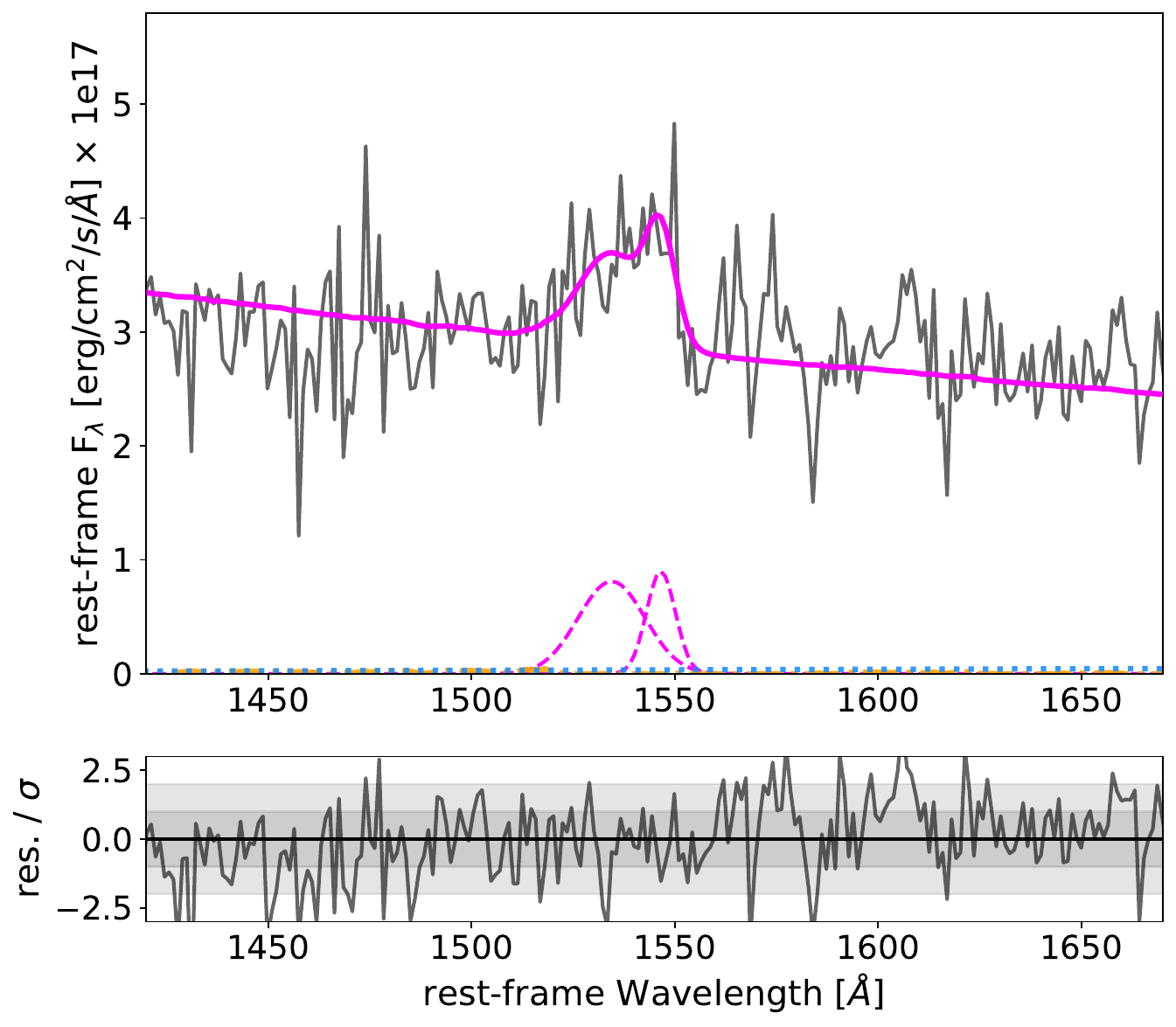}
	\includegraphics[width=0.559\hsize]{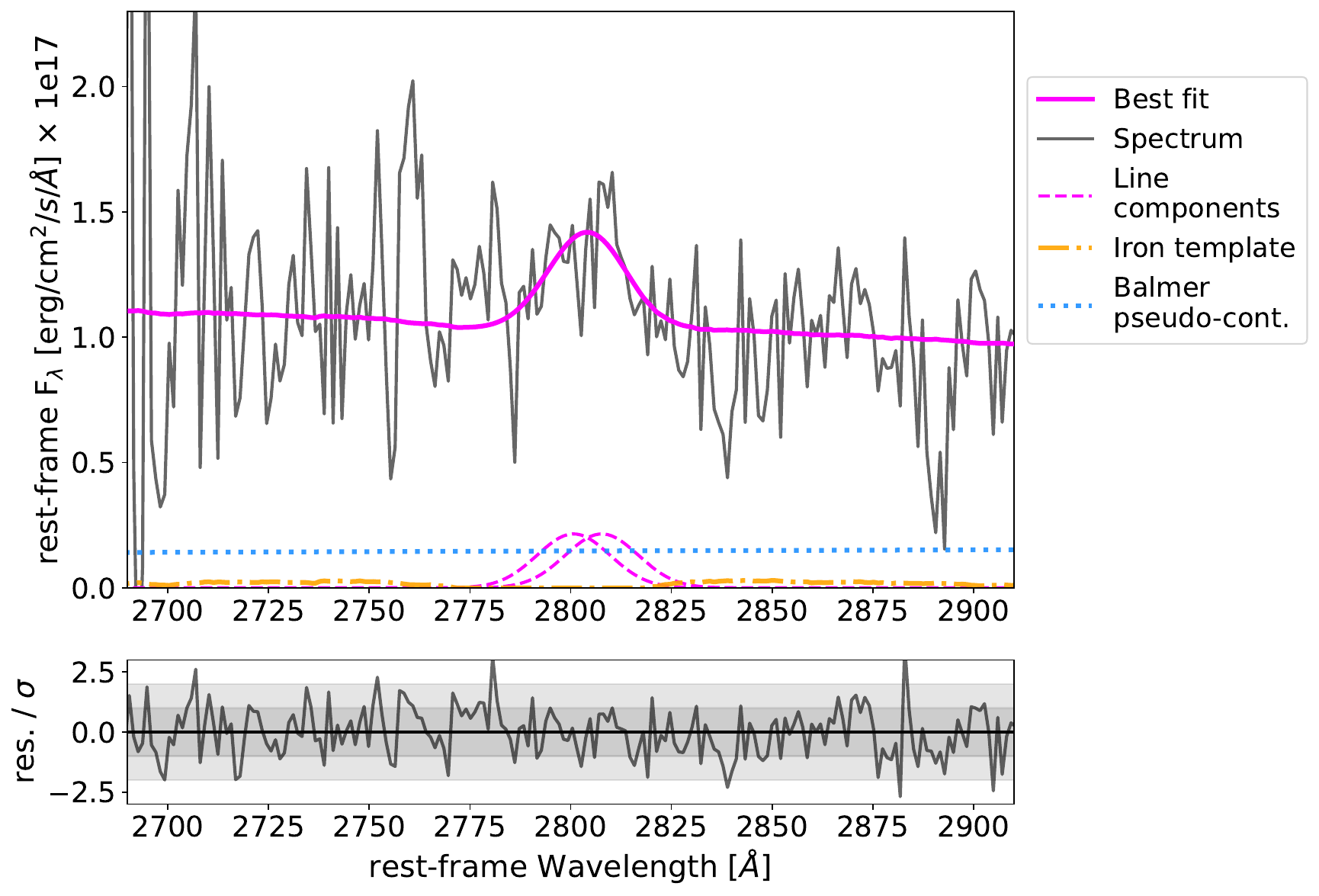}
    \caption{{\bf Top panels:} Zoom on the rest-frame CIV (left) and MgII (right) emission lines of VIK~J2318$-$31. The grey line shows the rest-frame spectrum. The best-fit, continuum plus emission lines, is shown as a solid magenta line. In both cases we used two Gaussian components (dashed magenta lines), but in the case of the MgII line we fixed the width and the normalization of the components to be equal and the relative position of the peaks to 7.2~\AA. The Balmer and the iron pseudo-continuum contributions are shown as a dotted blue line and a dashed-dotted orange line, respectively. {\bf Bottom panels:} residual of the fit divided by the error on the spectrum at each wavelength. Shaded areas indicate $\pm$1,2$\sigma$.}
    \label{fig:VIK_lines}
\end{figure*}

In order to analyse the broad emission lines seen in VIK~J2318$-$31, we started by considering the rest-frame spectrum assuming a redshift $z$=6.4429$\pm$0.0003 \citep[see][]{Venemans2020}.
Given the low signal-to-noise (S/N~$\sim$10), we performed a fit to the CIV emission line using two Gaussian functions, as opposed to more components, often considered in other works \citep[e.g.,][]{Coatman2017}. We report in Fig. \ref{fig:VIK_lines} a zoom-in on the CIV line together with the best fit components and we report in Tab. \ref{tab:CIV_line} the best-fit parameters obtained. 
Similarly to \cite{Diana2021}, the uncertainties were computed using a Monte Carlo approach, where we simulated 1000 independent spectra based on the best-fit continuum and line with a Gaussian-distributed noise added (derived from the RMS of the continuum around the line). We then applied the fit to each mock spectrum to determine the best-fit parameters. Finally, we used the standard deviation of each parameter distribution as its uncertainty.

As shown in Fig. \ref{fig:VIK_lines}, the CIV line profile can be described by two Gaussian functions, a blue-shifted broad component with $\lambda_{\rm cen}=1533.5$~\AA \, and $\sigma=7.7$~\AA \,  and a narrower component with $\lambda_{\rm cen}=1546.1$~\AA \, and $\sigma=3.6$~\AA \, (strictly speaking, this is still a broad emission line, having FWHM=1880~km~sec$^{-1}$). 
In order to compute the total blue shift we used $\Delta v$ = $c$(1549.5\AA \, $-$\,$\lambda_{\rm half}$)/1549.5\AA \, \citep[e.g.,][]{Coatman2017}, where 1549.5\AA \, is the rest-frame wavelength of the CIV line, $c$ is the speed of light and $\lambda_{\rm half}=1539\pm2$ is the line centroid, that is, the bisector of the cumulative line flux. 
In the case of VIK~J2318$-$31 we find $\Delta v=2110$~km sec$^{-1}$, which is a relatively large value, but it is still consistent to what is typically measured in high-redshift QSOs \citep[see, e.g., tab. 1 in ][]{Mazzucchelli2023}. 
The presence of a large CIV blue shift normally indicates that the profile of the line is strongly affected by emission from regions in non-virial motions, that is, with strong outflows (e.g., \citealt{Baskin2005}). 
Interestingly, while RL QSOs present, on average, smaller blue shifts in the CIV with respect to RQ QSOs \citep[see, e.g.,][]{Richards2011}, the value derived for VIK~J2318$-$31 is the largest currently found in $z>6$ RL QSOs (see, e.g., \citealt{Banados2021,Belladitta2022}), potentially indicating the presence of important outflows, and, consequently, of feedback (e.g., \citealt{Maiolino2012}), also for the RL population in the early Universe.
At the same time, given the rest-frame equivalent width (REW) of the CIV line (8.8$\pm$0.9~\AA), the presence of a large blue-shift is consistent with the trend of increasing blue-shift values for decreasing REW \citep[e.g.][]{Vietri2018}. 

Given the asymmetry due to the blue-shifted component observed in the CIV line profile of VIK~J2318$-$31, in order to estimate the BH mass we considered the empirical correction derived by \cite{Coatman2017} to the FWHM of the CIV line (eq. 4 in their paper), which was computed from the comparison of the results obtained with the CIV and the H$_{\alpha}/$H$_{\beta}$ emission lines. In this way, the relation used to estimate the BH mass should have a significantly smaller scatter ($\sim$0.2~dex) compared to the un-corrected values. Using the following empirical relation, derived by \cite{Vestergaard2006}, we computed the mass of the central BH of VIK~J2318$-$31:

\begin{equation}
 \begin{aligned}
    {\rm M_{\rm BH}} = 10^{6.66} \times \left( \frac{\rm FWHM_{corr.}(CIV)}{\rm 10^3~km~sec^{-1}}\right)^2 \times \\
    \times \left(\frac{\lambda L_{1350\text{\normalfont\AA}}}{\rm 10^{44}~erg~sec^{-1}} \right)^{0.53} {\rm M_\odot},
\end{aligned}
    \label{eq:BHM_civ}
\end{equation}
where the $\lambda L_{1350\text{\normalfont\AA}}$ luminosity was computed directly from the spectrum continuum. Using this relation, we find $\textrm{M}_{\rm BH}=8.1\pm5.0\times10^{8}~{\rm M_\odot}$, where the statistical errors are comparable to the scatter of the scaling relation: ${\sim} 0.2$~dex; with the blue-shift correction, \cite{Coatman2017}, or ${\sim} 0.36$~dex without this correction, \cite{Vestergaard2006}.

At the same time, we also used the MgII line in order to compute the SMBH mass. 
We followed a similar procedure for the analysis of the MgII line and the computation of the related uncertainties, with respect to the CIV line. We considered again two Gaussian components, since the MgII line is composed by a doublet with central wavelengths 2795.5\AA \, and 2802.7\AA. However, in this case, we used two Gaussian functions with the same width and normalisation and with the relative peak wavelength separation fixed to 7.2\AA. In this way, we still had three free parameters, as in the case of a single Gaussian. We report a zoom on the MgII together with its best fit in Fig. \ref{tab:CIV_line}, right. The estimates, and the corresponding errors, obtained from the fit are reported in Tab. \ref{tab:CIV_line}. Uncertainties were computed, once again, with Monte Carlo simulations.
In order to derive the BH mass of VIK~J2318$-$31 from the MgII line we used the following scaling relation (from \citealt{Vestergaard2009}):
\begin{equation}
 \begin{aligned}
    {\rm M_{\rm BH}} = 10^{6.86} \times \left( \frac{\rm FWHM(MgII)}{\rm 10^3~km~sec^{-1}}\right)^2 \times \\
    \times \left(\frac{\lambda L_{3000\text{\normalfont\AA}}}{\rm 10^{44}~erg~sec^{-1}} \right)^{0.5} {\rm M_\odot},
\end{aligned}
    \label{eq:BHM_mgii}
\end{equation}
where the $\lambda L_{3000\text{\normalfont\AA}}$ luminosity was computed directly from the spectrum continuum. Using this relation, we find ${\rm M}_{\rm BH}=4.2\pm3.0\times10^{8}~{\rm M_\odot}$, where the scatter of the relation is ${\sim} 0.55$\,dex.

\begin{table*}
	\centering
	\caption{Results of the fit of the CIV and MgII broad emission lines in VIK~J2318$-$31. The uncertainties of the mass and accretion rate reported in this table do not take the scatter of the scaling relations used for their derivation into account.}
	\label{tab:CIV_line}
\begin{threeparttable}
\centering

\begin{tabular}{cccccccccc}

\hline
\hline
FWHM  & FWHM$_{\rm cor}$   & L$_{\rm cont.}$ & L$_{\rm line}$ & $\lambda_{\rm half}$ & $\Delta v$ & M$_{\rm BH}$ & $\lambda_{\rm Edd}$\\
(km sec$^{-1}$)  & (km sec$^{-1}$) & (erg sec$^{-1}$) & (erg sec$^{-1}$) & \AA & (km sec$^{-1}$) & M$_\odot$ \\
\hline
CIV at 1549\AA &&&L$_{\rm 1350\text{\normalfont\AA}}$&&\\
4740$\pm$1150 & 3190$\pm$980 &  2.2$\pm0.1\times$10$^{46}$  & 1.2$\pm$0.2$\times$10$^{44}$ & 1539$\pm$2 & 2110$\pm$280 & 8.1$\pm$5.0$\times$10$^8$ & 0.8$\pm$0.6\\
\hline
MgII at 2798\AA &&&L$_\mathrm{3000\text{\normalfont\AA}}$&&\\
2270$\pm$810 & --  & 1.3$\pm0.1\times$10$^{46}$ &  4.6$\pm$0.9$\times$10$^{43}$ & 2804$\pm$3 & $-$670$\pm$290 & 4.2$\pm$3.0$\times$10$^8$ & 1.2$\pm$0.9\\

\hline
 \end{tabular}
\end{threeparttable}
\end{table*}


\subsubsection{Bolometric luminosity and Eddington ratio}

Having an estimate for the mass of the SMBH in VIK~J2318$-$31, we can now derive how fast the SMBH is accreting. In particular, following the literature, we can compute the $\lambda_{\rm EDD}$ parameter, defined as the ratio between the bolometric luminosity (L$_{\rm bol}$), and the Eddington Luminosity L$_{\rm EDD}$. 
To estimate L$_{\rm bol}$ we used the continuum luminosity of VIK~J2318$-$31 derived from its NIR spectrum (L$_{1350\text{\normalfont\AA}}=2.2\pm0.1\times10^{46}$~erg~sec$^{-1}$) and the bolometric corrections computed by \cite{Shen2008} ($\rm K = 3.81 \pm 1.26$). In this way we find: L$_{\rm bol}$ = K $\times$ L$_{1350\text{\normalfont\AA}} = 8.4\pm2.8\times 10^{46}$~erg~sec$^{-1}$ and $\lambda_{\rm EDD} = 0.8 \pm 0.6$. If we consider the MgII estimate and the continuum at 3000\AA, we obtain: L$_{\rm bol} = 6.6\pm1.6\times 10^{46}$~erg~sec$^{-1}$ and $\lambda_{\rm EDD} = 1.2 \pm 0.8$. Interestingly, in both cases, observations suggest that VIK~J2318$-$31 is hosting a highly accreting SMBH, of the order of its Eddington limit.

\subsubsection{Accretion disc model}
In order to test the presence of possible biases in the BH mass estimate from broad emission lines, we also used another independent method. In particular, we modelled the continuum optical-NIR (UV-optical in the rest frame) emission of VIK~J2318$-$31 with an accretion disc (AC) model\footnote{See \url{https://github.com/FabioRigamonti/pyADfit.git} for the code used in this work.}  \citep[e.g.,][]{Calderone2013,Ghisellini2015b,Diana2021}.
Following the \citealt{Shakura1973} (SS) model, we assumed that the optical/UV continuum emission of the AGN is produced by an optically thick, geometrically thin accretion disk (AD) composed by rings that emit as a black body (BB) with different temperatures, where the inner radius of the disc is set to $\rm R_{\rm in}=3R_{\rm s}$ and the outer radius to R$_{\rm out} = 10^4$R$_{\rm s}$\footnote{where $\rm R_{\rm s}=2GM_{\rm BH}/c$ is the Schwarzschild radius of the BH.}. In this model, the free parameters available are the BH mass (${\rm M_{\rm BH}}$), the accretion rate ($\dot{M}$) and the viewing angle ($\theta$). We fixed the latter to $\theta=30^{\rm o}$, since, as mentioned before, this source is an un-obscured AGN with a jet that is not oriented close to our line of sight. Moreover, we note that, we are technically considering a non-rotating SMBH, even though VIK~J2318$-$31 is a RL QSO and therefore expected to be associated with a spinning BHs \citep[e.g.][]{Tchekhovskoy2010}. However, as in similar studies (e.g., \citealt{Belladitta2022,Sbarrato2022}), this choice can be justified by the fact that a SS accretion disc with $\rm R_{\rm in}=3R_{\rm s}$ is a good approximation of a rotating BH with spin $a=0.71$ (based on the KERRBB model discussed in \citealt{Li2005}; see, e.g., fig. A2 in \citealt{Calderone2013}). 
Under these assumptions, the overall disc luminosity is given by:
\begin{equation}
    {\rm L}(\nu, {\rm M}_{\rm BH}, \dot{\rm M}){\rm d}\nu = \int^{\rm R_{\rm out}}_{\rm R_{\rm in}} {\rm r B_{\nu}}[{\rm T(r,M_{\rm BH},}\dot{\rm M})]{\rm d\nu dr},
    \label{eq:lum_disc}
\end{equation}
where r is the distance from the centre of the BH and B$_\nu$ is the Planck's BB spectral density at the temperature T. The temperature of the BB depends on the distance from the BH (as T $\propto$ r$^{-3/4}$), the mass of the BH (as T $\propto$ M$_{\rm BH}^{-1/2}$) and the accretion rate (as T $\propto$ r$^{1/4}$). In particular, the higher the mass of the BH is, the lower the frequency peak of the SED is. 

\begin{figure}
\centering
	\includegraphics[width=0.85\columnwidth]{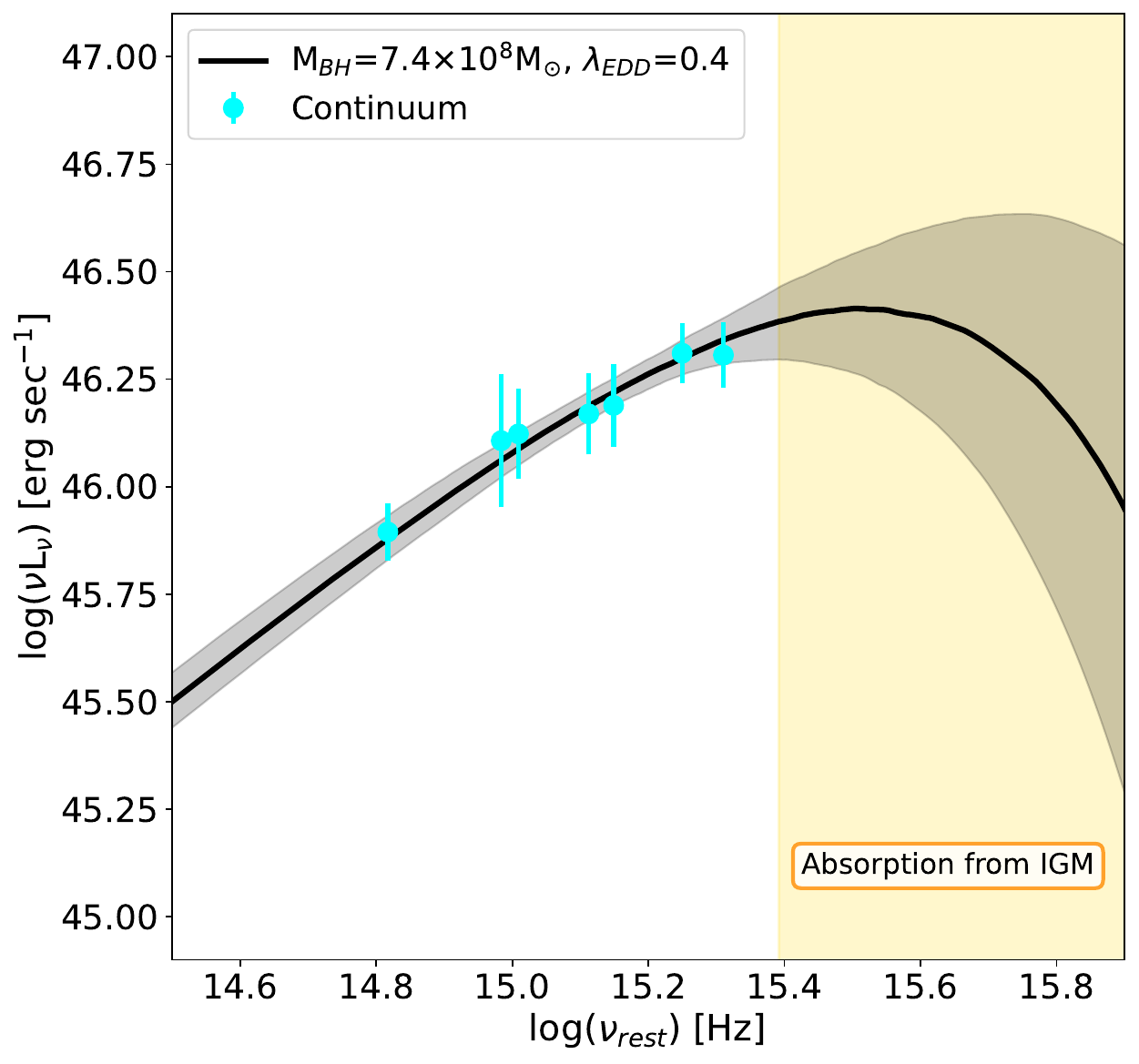}
	\includegraphics[width=\columnwidth]{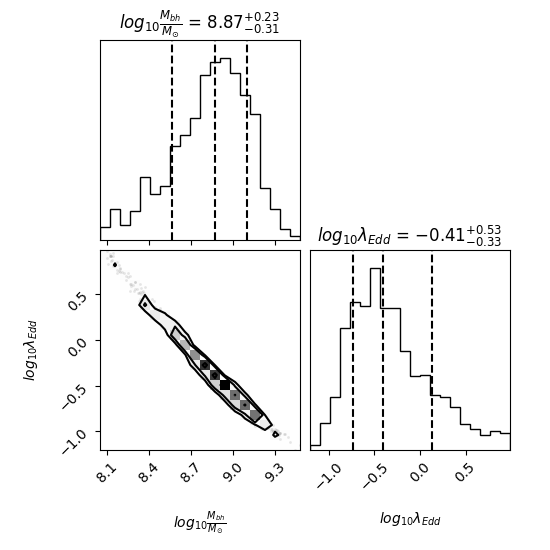}
    \caption{{\bf Top panels:} Rest-frame optical-UV luminosities of VIK~J2318$-$31 (cyan points) together with the best-fit SS AD model (solid black line). The shaded blue area represents the model within the 16$^{\rm th}$ and 84$^{\rm th}$ percentiles. The hatched yellow region at $\nu>3.3\times10^{15}$~Hz indicates frequencies higher than the Ly$\alpha$ emission line, which not accessible at high redshift due to the absorption from the IGM. {\bf Bottom panels:} Corner plot with the posterior distributions of the parameters used in the fit. Dashed vertical lines indicate the 16$^{\rm th}$, 50$^{\rm th}$ and 84$^{\rm th}$ percentiles.}
    \label{fig:VIK_ADmod}
\end{figure}

In order to perform a fit to the model outlined above, we considered the continuum emission from the observed NIR spectrum (GNIRS+X-SHOOTER) described in Sec. \ref{sec:data}. In particular, we averaged the NIR spectrum in ranges unaffected by telluric adsorptions and without emission lines (see cyan shaded regions in Fig. \ref{fig:VIK_NIR_spec}). As a conservative estimate of the mean flux density errors, we considered the RMS of the spectrum (smoothed to a common 500~km~sec$^{-1}$ resolution), by noting that these errors do not significantly impacting the result of the fit, since they have similar values. We report in Tab. \ref{tab:photo_spec} the magnitudes and the corresponding errors obtained. Moreover, in the fit, we also consider the {\it W1} band (mag = 20.71$\pm$0.13; in the AB system) from the {\it Wide-field Infrared Survey Explorer} \citep{Wright2010} catalogue (catWISE; \citealt{Eisenhardt2020}). From the fit to these data-points, we obtain M$_{\rm BH}=7.4^{+5.2}_{-3.8}\times10^8~{\rm M_{\odot}}$ and $\lambda_{\rm EDD} = 0.4^{+0.5}_{-0.3}$. Fig. \ref{fig:VIK_ADmod} shows the best-fit model (solid black line) with the associated 1$\sigma$ uncertainty (shaded grey area) in the top panel and the corner plot of the parameters derived from the fit in the bottom panel.\\

Both M$_{\rm BH}$ and $\lambda_{\rm EDD}$ derived from the AD model are consistent with the values found through the SE estimate. This suggests that the values estimated from the CIV and MgII lines can be considered reliable within the extent of their uncertainty. Given the higher S/N\footnote{Estimated as the ratio between the REW of the line and its uncertainty.} of the CIV line ($\sim$10) compared to the MgII one ($\sim$5), we consider M$_{\rm BH}=8.1^{+6.8}_{-5.6} \times 10^8 {\rm M_{\odot}}$ and $\lambda_{\rm EDD} = 0.8^{+0.8}_{-0.6}$ as the best estimate for the mass and the accretion rate, respectively, of the central BH in VIK~J2318$-$31. Where we also added in quadrature the scatter of the scaling relation used in eq. \ref{eq:BHM_civ}).

\begin{table}
	\centering
	\caption{Photometric points computed from the continuum of the GNIRS+X-Shooter spectrum of VIK~J2318$-$31 binned at a resolution of 500~km~sec$^{-1}$ (see Fig. \ref{fig:VIK_NIR_spec} for the wavelength range considered). The errors reported are the RMS of the spectrum in the range used to compute the photometric points.}
	\label{tab:photo_spec}
\begin{threeparttable}
\centering
\begin{tabular}{lcccccc}

\hline
\hline
$\lambda_{\rm cen.}$   &  mag  \\
(\AA) & (AB) \\
\hline
 10913 & 20.93$\pm$0.15 \\
 12551& 20.75$\pm$0.14 \\
 15825 & 20.80$\pm$0.18 \\
 17216 & 20.76$\pm$0.17 \\
 21801 & 20.61$\pm$0.19 \\
 23029 & 20.61$\pm$0.25 \\
\hline
\hline

 \end{tabular}
\end{threeparttable}
\end{table}

\section{Discussion and Conclusions}
\label{sec:conclusions}

As described in the previous sections, VIK~J2318$-$31 hosts a $\sim$8$\times 10^8 {\rm M_{\odot}}$ SMBH accreting close to the Eddington limit. The radio luminosity and spectral shape of this source suggests that the radio emission is powered by young relativistic jets (see also the discussion in \citealt{Ighina2022b}). Indeed, the compact radio morphology \citep{Zhang2022} together with the presence of a turnover around $\sim$650~MHz (observed frame) in the radio spectrum are typical sings of young radio jets (see, e.g., \citealt{Odea2021}).
Even though it is not possible to confirm the origin of the turnover in the radio spectrum of VIK~J2318$-$31, due to the lack of information at very low frequencies ($\lesssim200$~MHz), it seems unlikely to be generated by SSA, since it would require an emitting region orders of magnitude smaller than what is typically found in radio peaked sources \citep[e.g.,][]{Orienti2008,Orienti2014}.
At the same time, the shape and the peak frequency derived for VIK~J2318$-$31 are consistent with previous studies on $z\gtrsim5$ RL QSOs \citep[e.g.,][]{Shao2020,Shao2022}.


From the X-ray point of view, the faint and steep ($\Gamma \gtrsim 1.7$) emission observed in VIK~J2318$-$31 suggests that relativistic boosting has only a marginal impact and, therefore, that the jets are oriented away from our line of sight. Interestingly, VIK~J2318$-$31 is the faintest $z>6$ RL QSO currently known \citep{Medvedev2020, Moretti2021}, even though the current number of such sources with dedicated X-ray observations is extremely limited.


With respect to other $z\gtrsim5.7$ bright QSOs with a SE estimate (either from the CIV or MgII line) from the literature, the properties of the SMBH hosted in VIK~J2318$-$31 are consistent with the distributions observed. In particular, compared to the samples analysed in \cite{Shen2019} and \cite{Mazzucchelli2023}, $\sim$90 individual sources in total, the BH mass derived for VIK~J2318$-$31 falls in the lower end of the distribution (within the lowest $\sim10$\%), while its accretion rate is fully consistent with what is typically found in these high-$z$ systems, that is, an accretion close to the Eddington limit (median value $\lambda_{\rm Edd}\sim0.7$). The agreement between the values obtained for VIK~J2318$-$31 and the rest of the high-$z$ population is likely due to the similar optical selection adopted for their discovery. Indeed, the recent discovery of many faint objects (about two orders of magnitude fainter in the optical compared to VIK~J2318$-$31) at $z>4$ with the JWST telescope revealed the presence of a large population of AGN (referred to as `little red dots'; see, e.g., \citealt{Matthee2023,Kocevski2023}) hosting $\sim10^{7-8}~{\rm M_\odot}$ SMBHs.

Even though the number of $z>5$ QSOs with dedicated NIR spectroscopy and analysis has significantly increased in the last few years \citep[e.g.,][]{Mazzucchelli2023,Lai2023}, only $\sim10$ of these sources are hosting relativistic jets (i.e., are radio loud; see \citealt{Yi2014,Banados2021,Diana2021,Belladitta2022,Belladitta2023}). Therefore, given the small statistics, it is not possible to derive solid conclusions on potential differences between the RL and RQ classes at these very high redshifts. 
While the mass and the accretion rate found in VIK~J2318$-$31 are similar to the other few high-$z$ SMBHs hosted in RL systems from the literature (again, due to a similar optical selection), this source presents one of the largest blue shift in the CIV line currently found in this specific QSO class (e.g., \citealt{Banados2021,Belladitta2022}). Given the weakness of the CIV line (REW~=~8.8$\pm$0.9~\AA), the large blue-shift value derived for VIK~J2318$-$31 is also consistent with the trend typically found in QSOs. Indeed, even though the origin of this relation is not clear yet \citep[see, e.g., discussion in][]{Sun2018}, the the faintest CIV lines typically present the largest blue shifts, with sources having REW[CIV]~$\lesssim$~15~\AA \, showing blue shifts up to $\sim4000-5000$~km~sec$^-1$ \citep[e.g.,][]{Ge2019}. 
Based on the threshold adopted in \cite{Diamond-Stanic2009}, VIK~J2318$-$31 is classified as a WELQs (i.e., REW[CIV]~$\lesssim$~15~\AA). Interestingly, recent studies seem to find a larger fraction of WELQs (based on a weak Ly$\alpha$ emission line) hosted in high-$z$ RL systems \citep[e.g.,][]{Gloudemans2022} compared to RQ QSOs at both high (e.g., \citealt{Banados2016}) and low redshift (\citealt{Diamond-Stanic2009}). 
The systematic spectroscopic study of these specific high-$z$ RL sources will help us in understanding the connection between relativistic jets, accretion and outflows in the early Universe.

The discovery of less massive \citep[e.g.,][]{Matthee2023} and/or more fast accreting \citep[e.g.,][]{Wolf2023} SMBHs at high redshift can help to alleviate the tension between current observations of massive SMBH hosted in $z\sim6.5$ QSOs and theoretical models describing their formation as seed BHs (see, e.g., fig. 10 in \citealt{Onoue2019}).
Indeed, the high accretion rate found in VIK~J2318$-$31 could explain the SMBH observed at $z=6.44$ as produced by a dense primordial star clusters with a seed mass of $\sim 10^4$~${\rm M_{\odot}}$ formed at $z\sim25$ and accreting with a radiative efficiency $\eta_{\rm d}\sim0.1$. However, the fact that this source is radio loud could imply that its radiative efficiency is closer to $\sim0.3$. In this case, an already massive BH ($\sim5\times10^7 {\rm M_{\odot}}$) is needed at $z\sim25$, which cannot be explained even by direct collapse of primordial gas clouds, since this can only happen after the first generation of stars ($z\sim10-15$; \citealt{Smith2019,Woods2019}). However, the very young age of the radio jet might suggest that the SMBH accreted with an efficiency $\eta_{\rm d}\sim0.1$ for most of its growth and only very recently spun up in order to produce and lunch the relativistic jets. Another potential way to explain the large mass observed could be that only a part of the gravitational energy of the accreting gas is heating up the disc \citep[e.g.,][]{Jolley2008,Jolley2009}, while the rest can be used to amplify the magnetic field, a necessary component for launching the relativistic jets \citep[e.g.,][]{Blandford1977}. Therefore, even though the total efficiency of the accreting process is still $\eta\sim0.3$, only a fraction ($\eta_{\rm d}\sim 0.1$; radiative efficiency; e.g., \citealt{Ghisellini2013,Ghisellini2015b}) is responsible for the disc luminosity and, as a consequence, of the Eddington limit. By having a larger Eddington limit to their accretion, SMBHs able to produce relativistic jets can grow faster in this scenario with respect to the canonical model and, potentially, even to the RQ counterparts (see, e.g., fig. 1 in \citealt{Connor2023}).
Finally, another possibility to ease the tension between the large masses observed and the little time available to grow, is the presence of multiple mergers in seed BHs. While, in general, at high redshift we do not expect mergers to be the dominant component in the mass growth of the seed BH population since they are too light to find a merging companion \citep[e.g.,][]{Volonteri2021}, from the study of the galaxy kinematics of VIK~J2318$-$31, \cite{Neeleman2021} found that the distribution of the [CII] emission can be potentially disturbed by a recent merger. Even though the results of \cite{Neeleman2021} are based on a low S/N image ($\sim$5), we cannot exclude that a fraction of the mass of the SMBHs hosted by VIK~J2318$-$31 comes from merger activity. 

As mentioned before, a statistical sample of RL QSOs at high redshift with a multi-wavelength coverage is needed to assess the importance of relativistic jets in the growth of the first seed BHs.

\begin{acknowledgements}
We want to thank the anonymous referee for their suggestions to improve the quality of the paper. We want to thank D. Dallacasa for his suggestions and comments regarding the radio properties of the source discussed here.\\
LI, AC and AM acknowledge financial support from the INAF under the project `QSO jets in the early Universe', Ricerca Fondamentale 2022.
NHW is supported by an Australian Research Council Future Fellowship (project number FT190100231) funded by the Australian Government.\\
This scientific work makes use of the Murchison Radio-astronomy Observatory, operated by CSIRO. We acknowledge again the Wajarri Yamatji people as the traditional owners of the Observatory site. Support for the operation of the MWA is provided by the Australian Government (NCRIS), under a contract to Curtin University administered by Astronomy Australia Limited. We acknowledge the Pawsey Supercomputing Centre which is supported by the Western Australian and Australian Governments.\\
The Australia Telescope Compact Array is part of the Australia Telescope National Facility (\url{https://ror.org/05qajvd42}) which is funded by the Australian Government for operation as a National Facility managed by CSIRO.
We acknowledge the Gomeroi people as the Traditional Owners of the Observatory site.\\
We thank the staff of the GMRT who have made these observations possible. The GMRT is run by the National Centre for Radio Astrophysics of the Tata Institute of Fundamental Research.\\
The scientific results reported in this article are based in part on observations made by the Chandra X-ray Observatory.\\
This work was enabled by observations made from the Gemini North telescope, located within the Maunakea Science Reserve and adjacent to the summit of Maunakea. We are grateful for the privilege of observing the Universe from a place that is unique in both its astronomical quality and its cultural significance.\\
Based on observations collected at the European Organisation for Astronomical Research in the Southern Hemisphere under ESO programme 097.B-1070.\\
This research made use of \texttt{PypeIt}, a Python package for semi-automated reduction of astronomical slit-based spectroscopy \citep{Prochaska2020a,Prochaska2020b}. This research made use of Astropy, a community-developed core Python package for Astronomy \citep{astropy2013,astropy2018}.
\end{acknowledgements}

%
%

\bibliographystyle{aa} 
\bibliography{referenze} 

\appendix

\section{Radio Images from MWA, uGMRT and ATCA}
\begin{figure}
\centering
	\includegraphics[width=0.85\hsize]{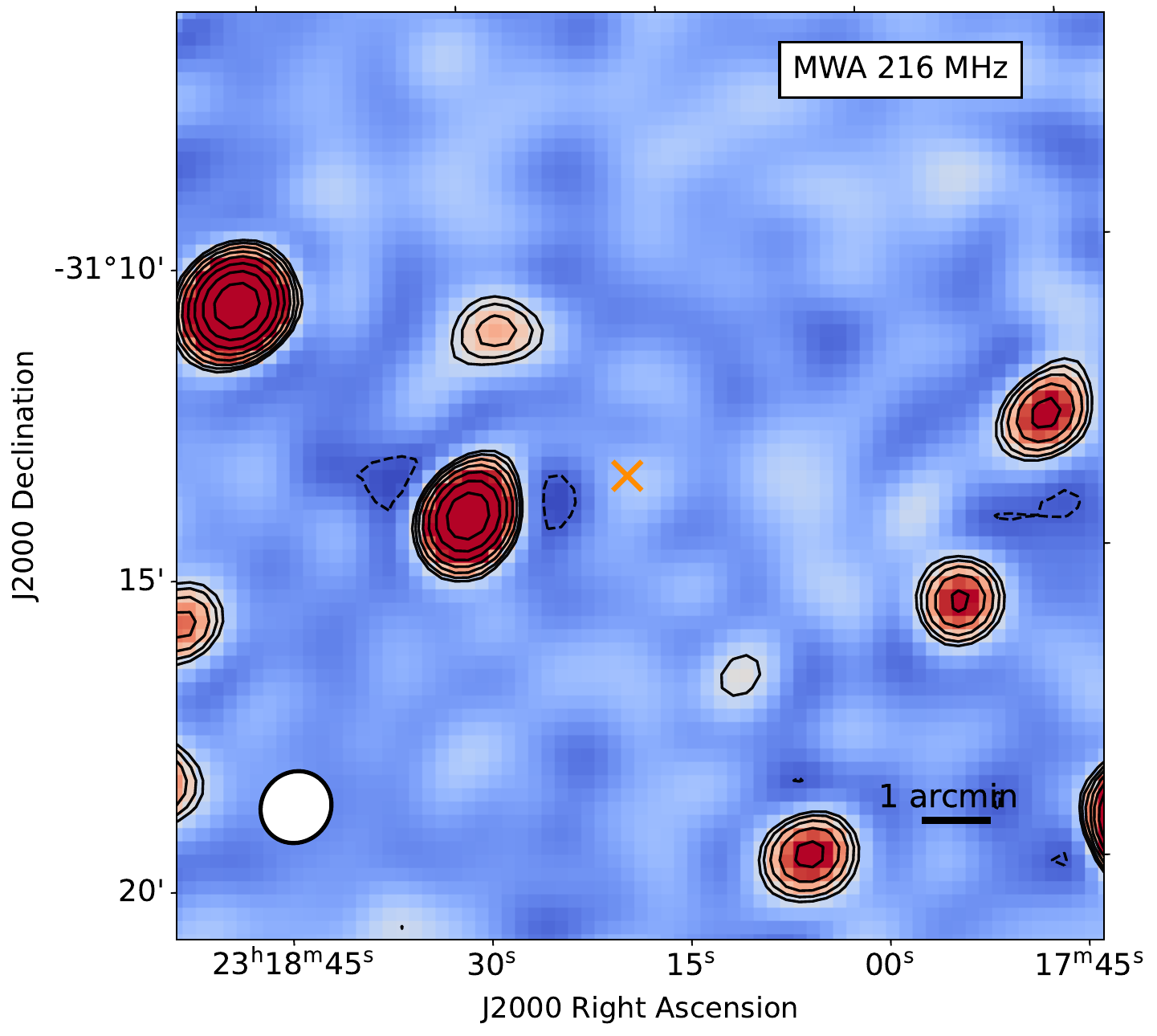}
    \caption{Cutout around the optical position of VIK~J2318$-$31 (orange cross) of the MWA observations of the G23 field. Contours start at $\pm$3$\times$RMS and increase by a factor of $\sqrt{2}$.}
    \label{fig:MWA}
\end{figure}

\begin{figure*}
\centering
	\includegraphics[width=0.33\hsize]{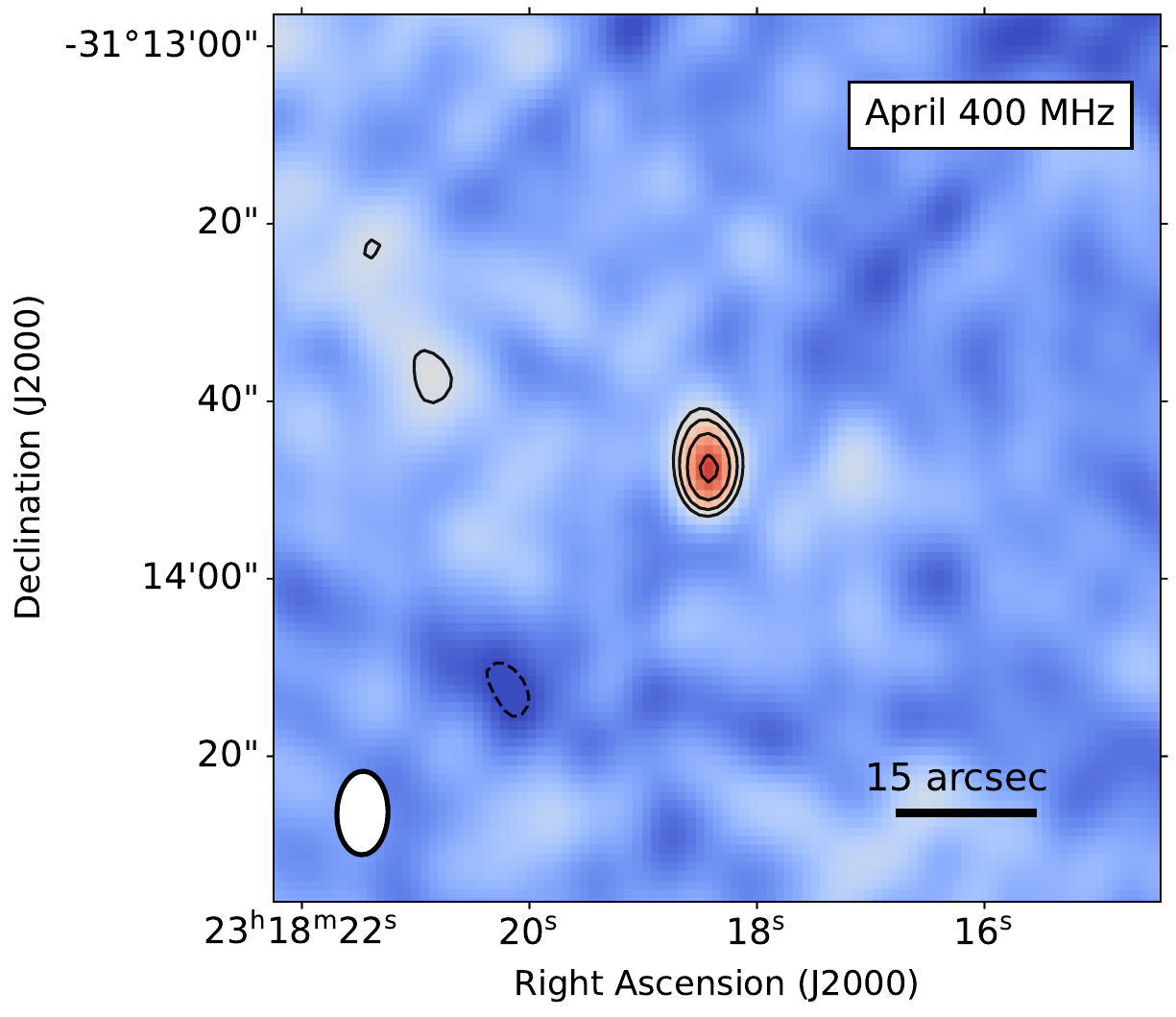}
	\includegraphics[width=0.33\hsize]{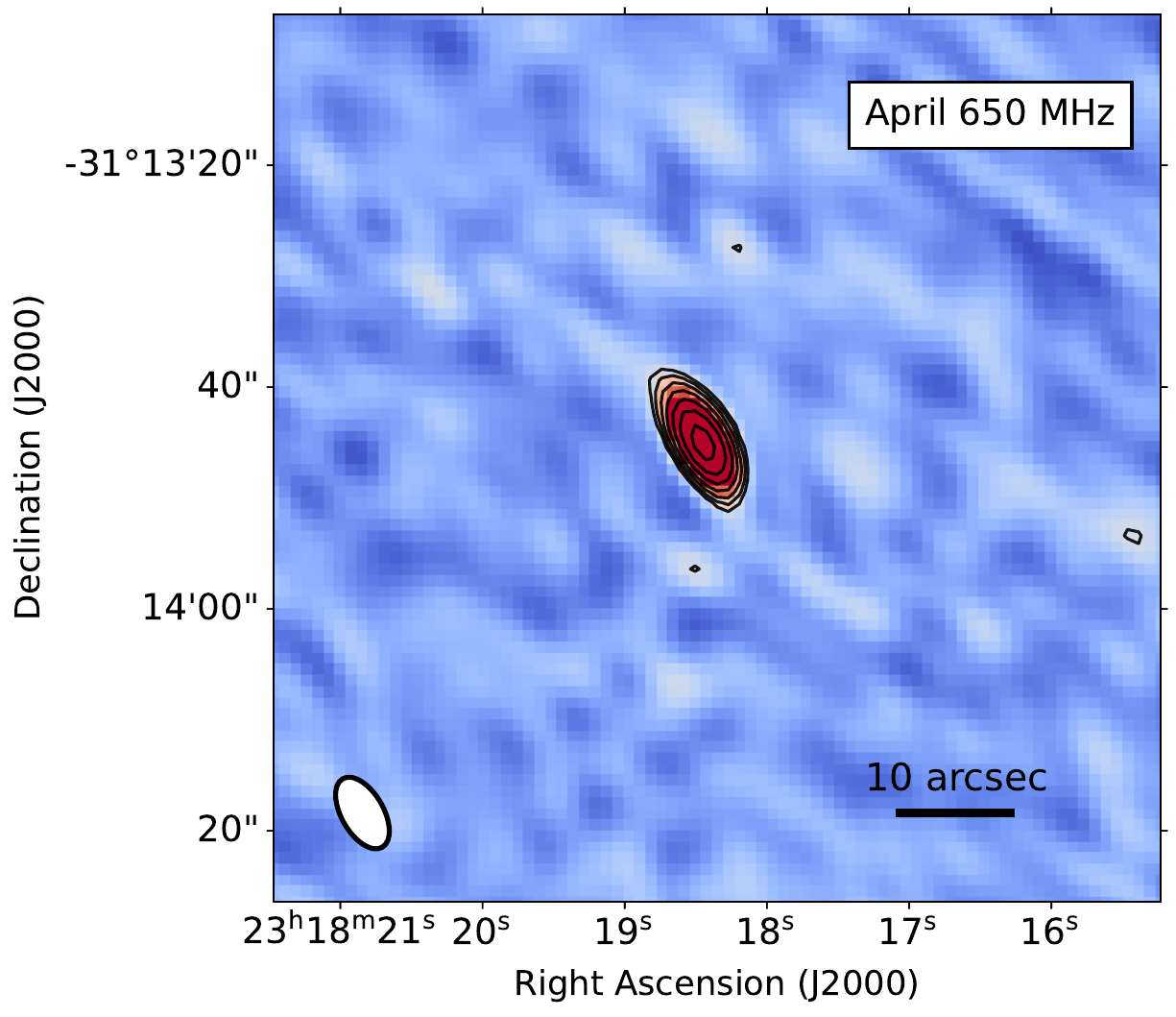}\\
 	\includegraphics[width=0.33\hsize]{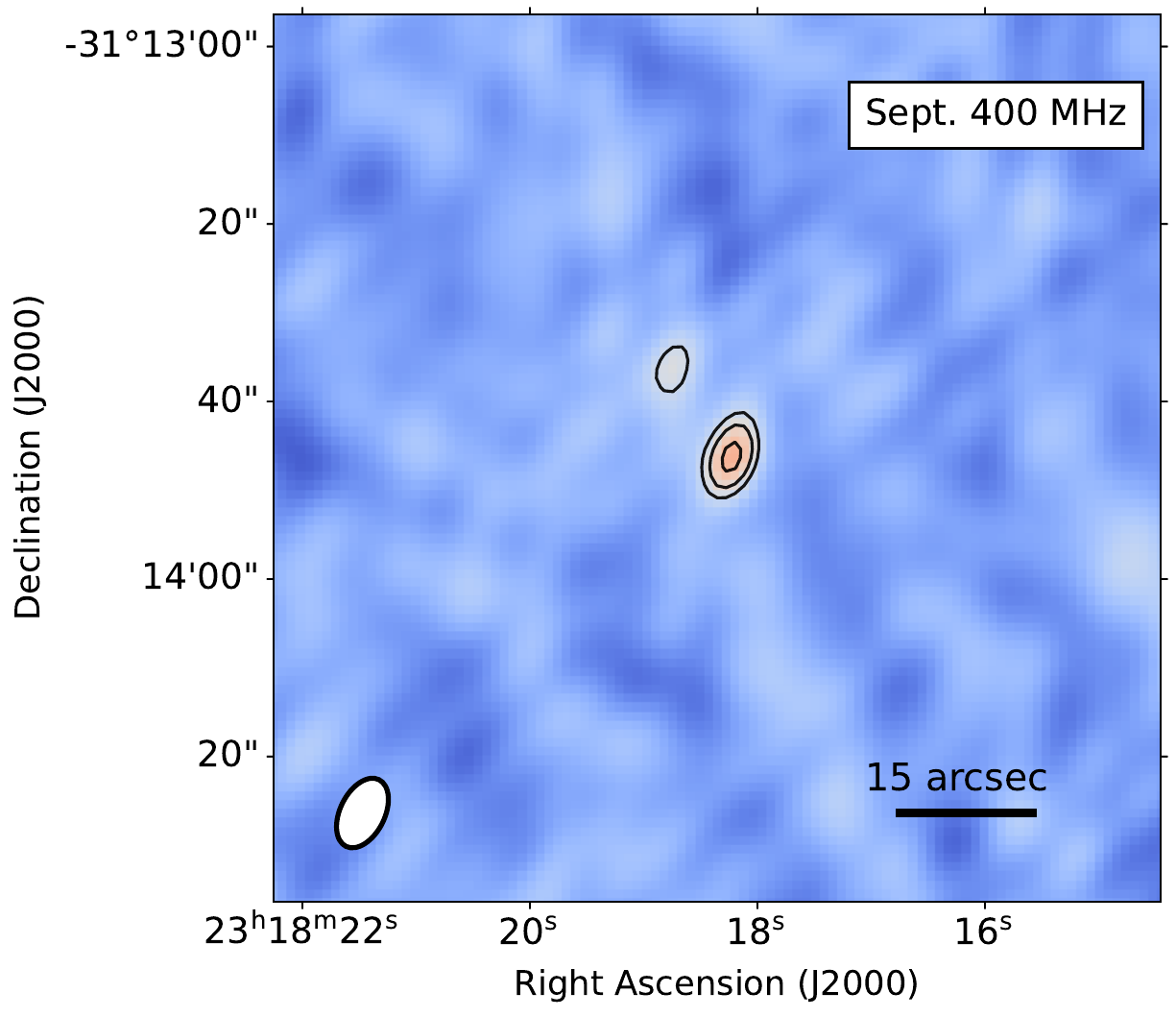}
	\includegraphics[width=0.33\hsize]{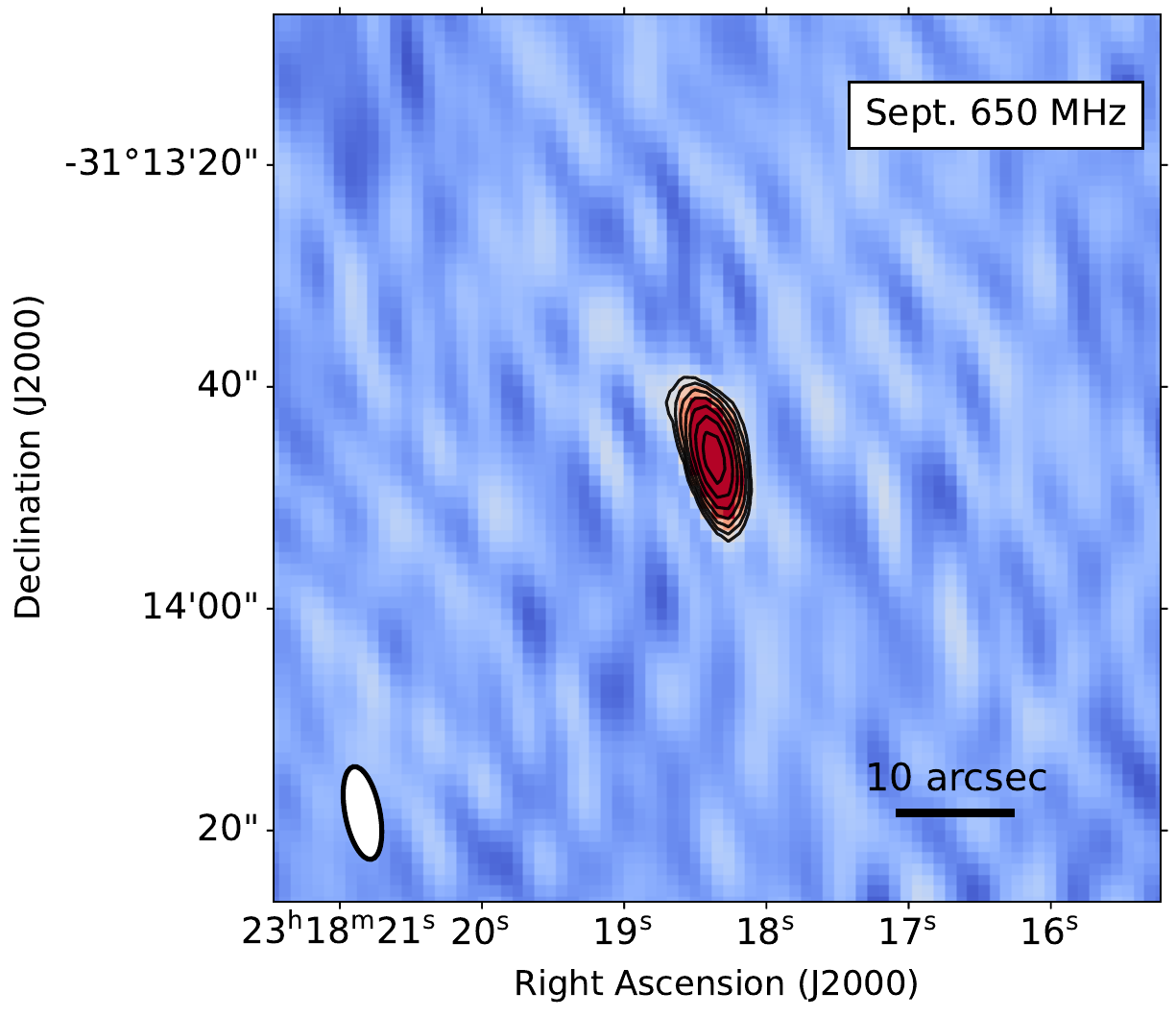}\\
    \caption{uGMRT observation of VIK~J2318$-$31 at 400 (left) and 650~MHz (right) in the two different epochs (April 2022, top, and September 2022, bottom). Contours start at $\pm$3$\times$RMS and increase by a factor of $\sqrt{2}$.}
    \label{fig:uGMRT}
\end{figure*}

\begin{figure*}
\centering
	\includegraphics[width=0.325\hsize]{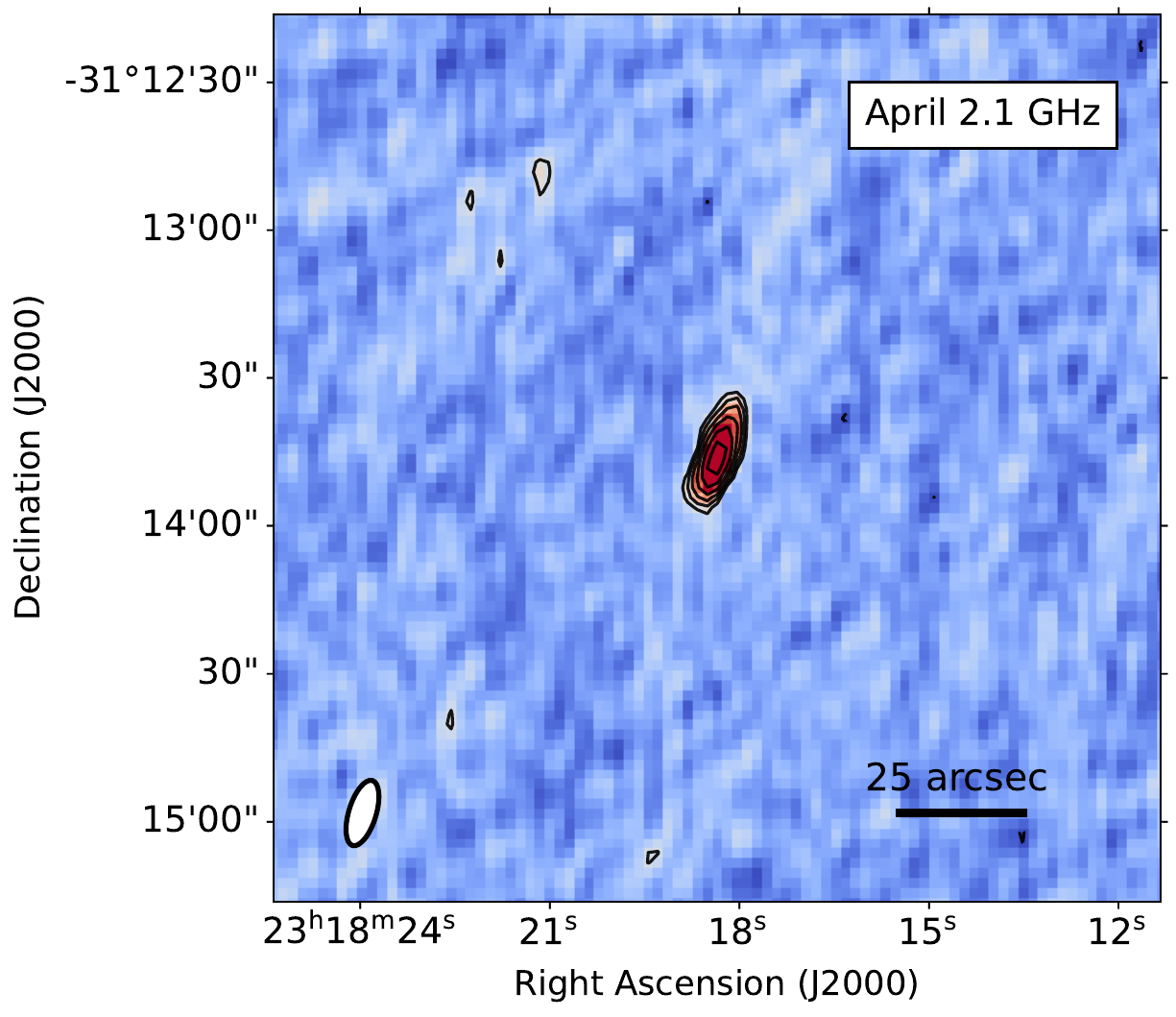}
	\includegraphics[width=0.325\hsize]{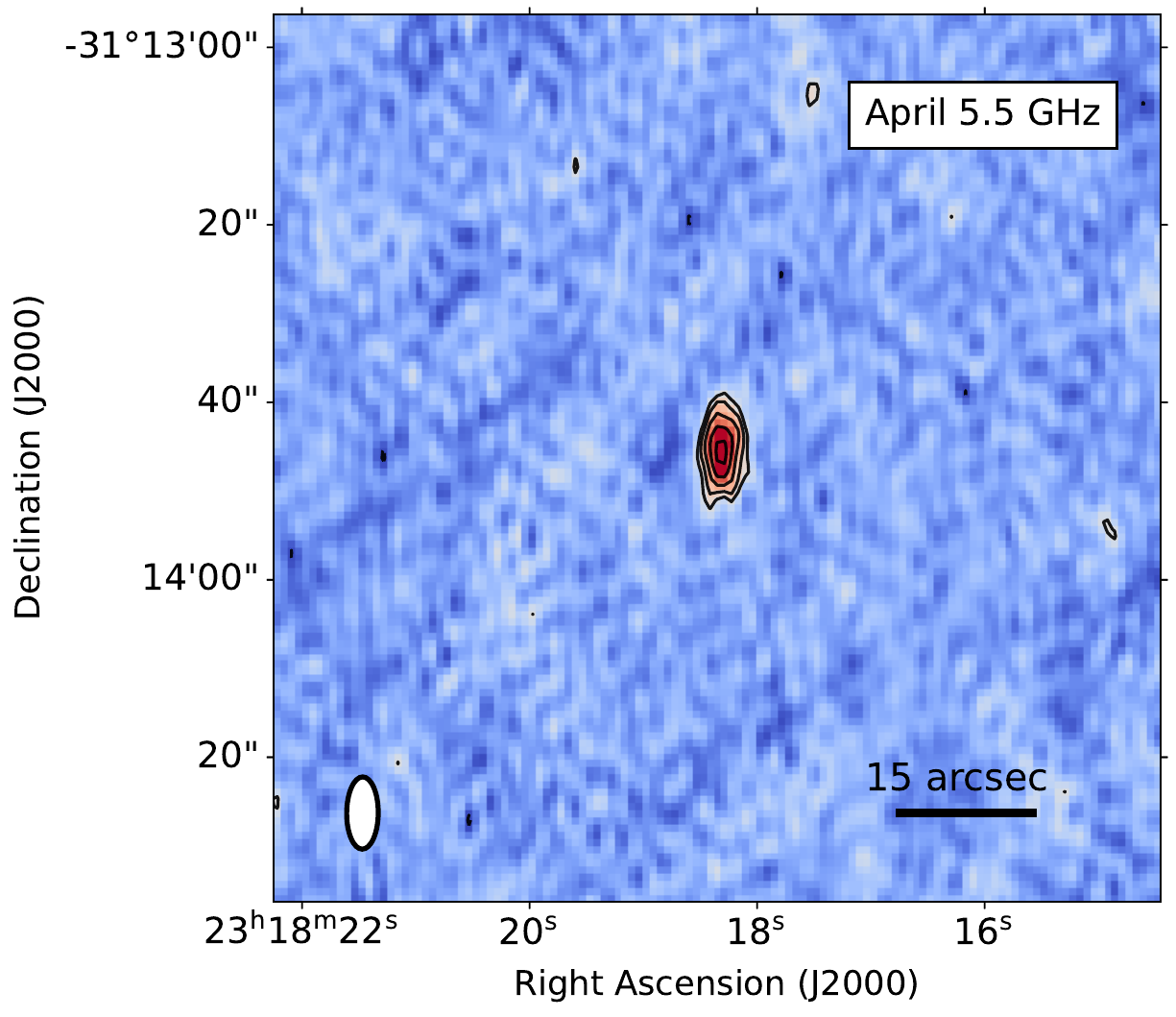}
	\includegraphics[width=0.325\hsize]{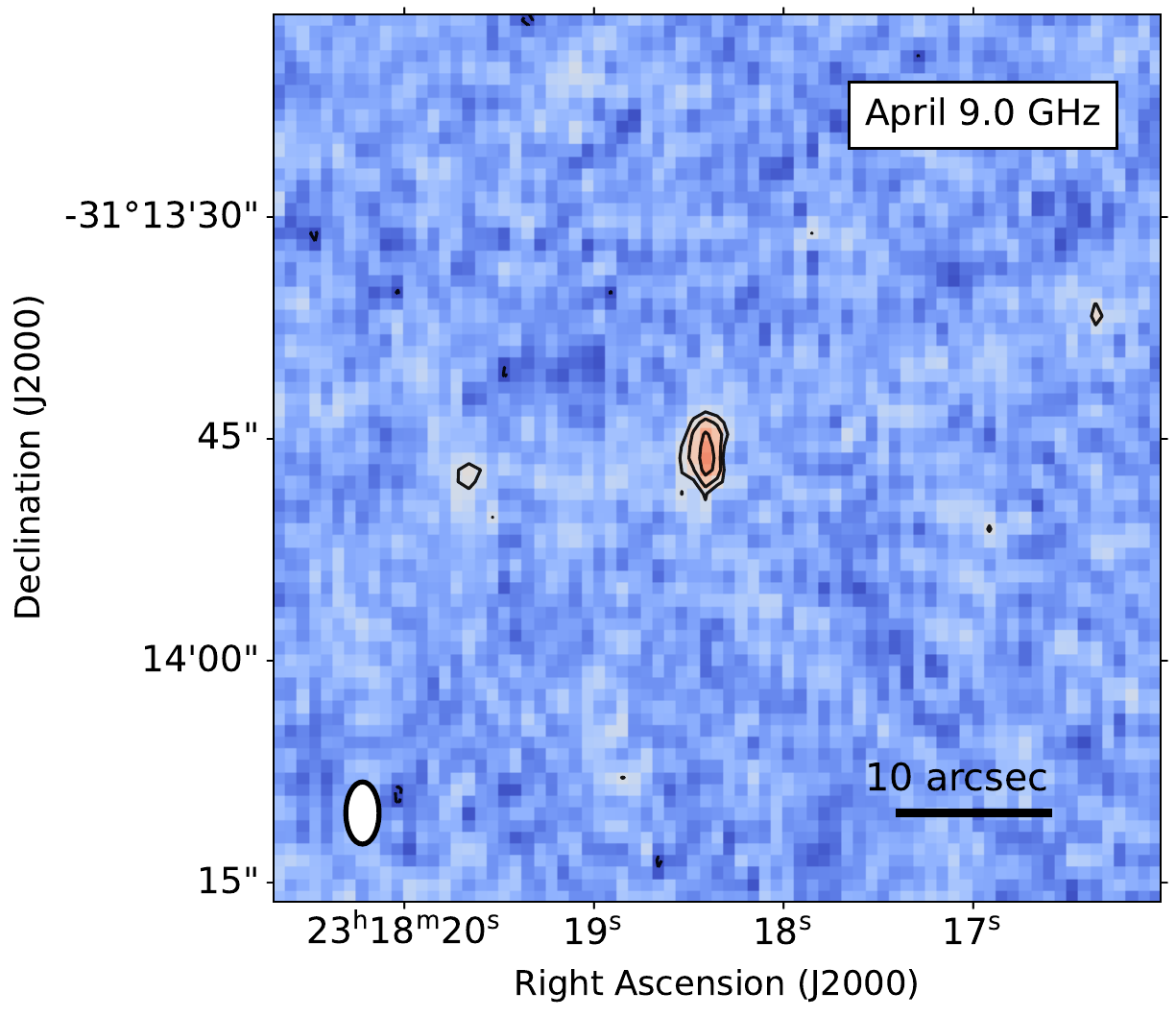}\\
	\includegraphics[width=0.325\hsize]{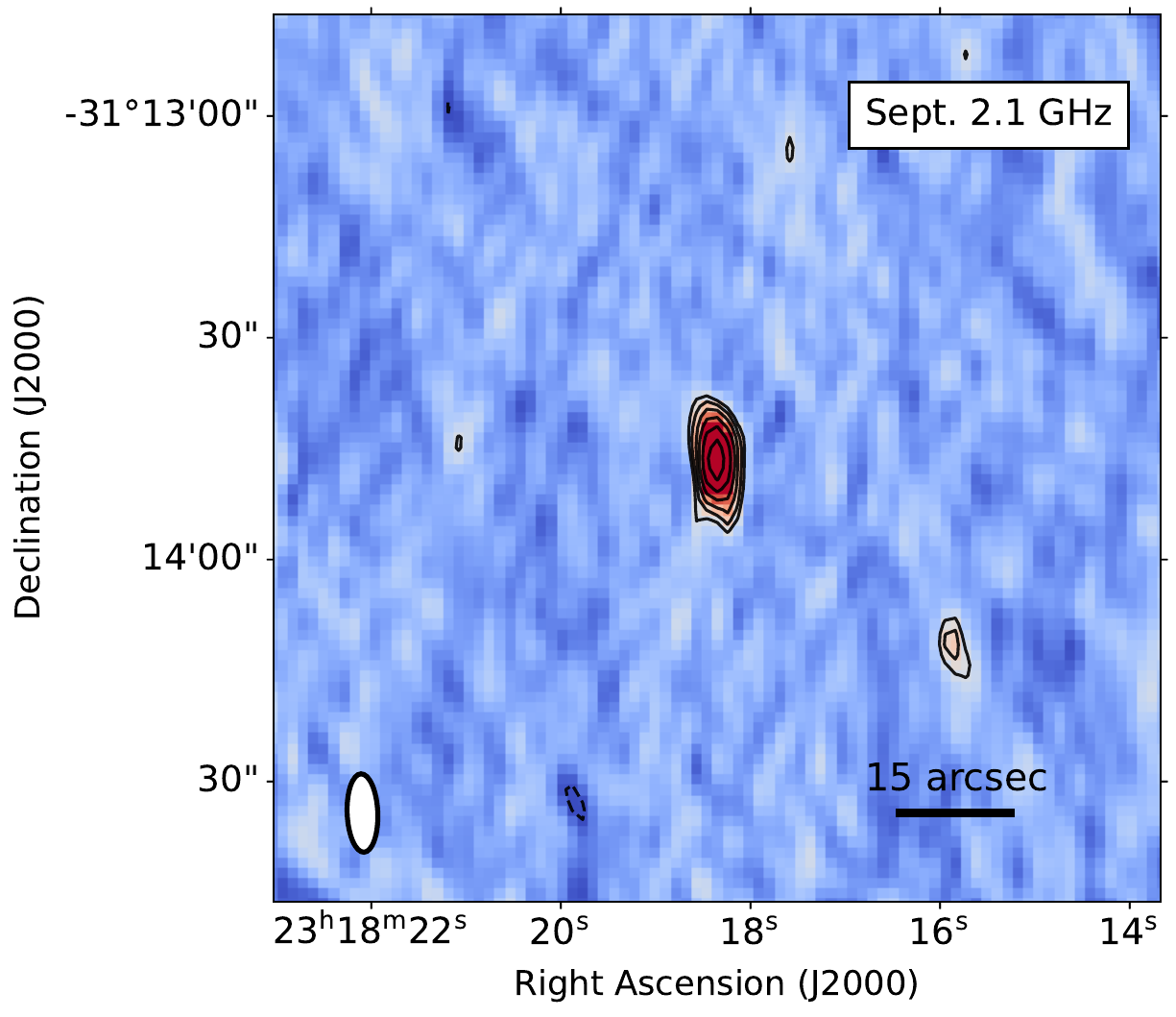}
	\includegraphics[width=0.325\hsize]{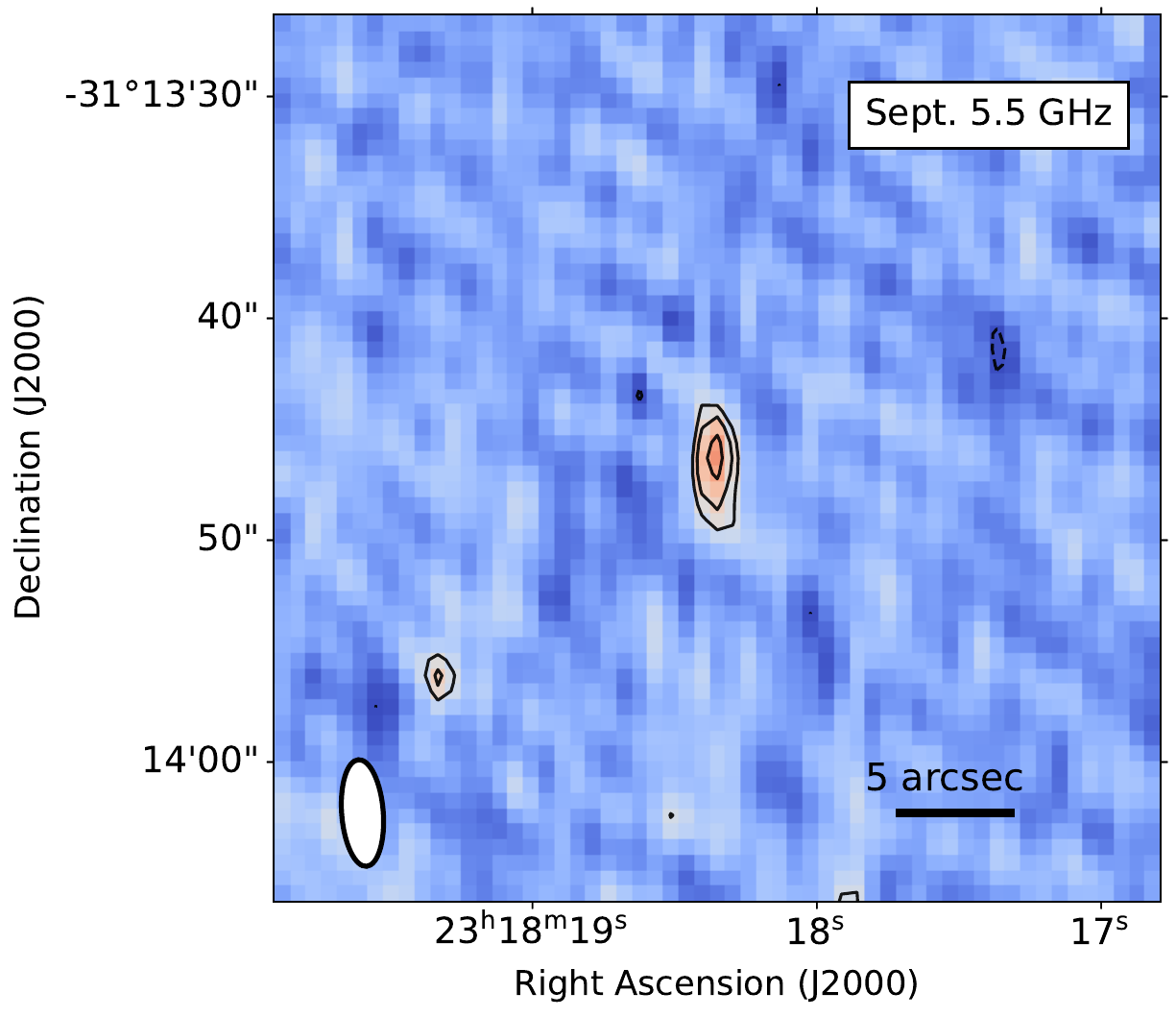}
	\includegraphics[width=0.325\hsize]{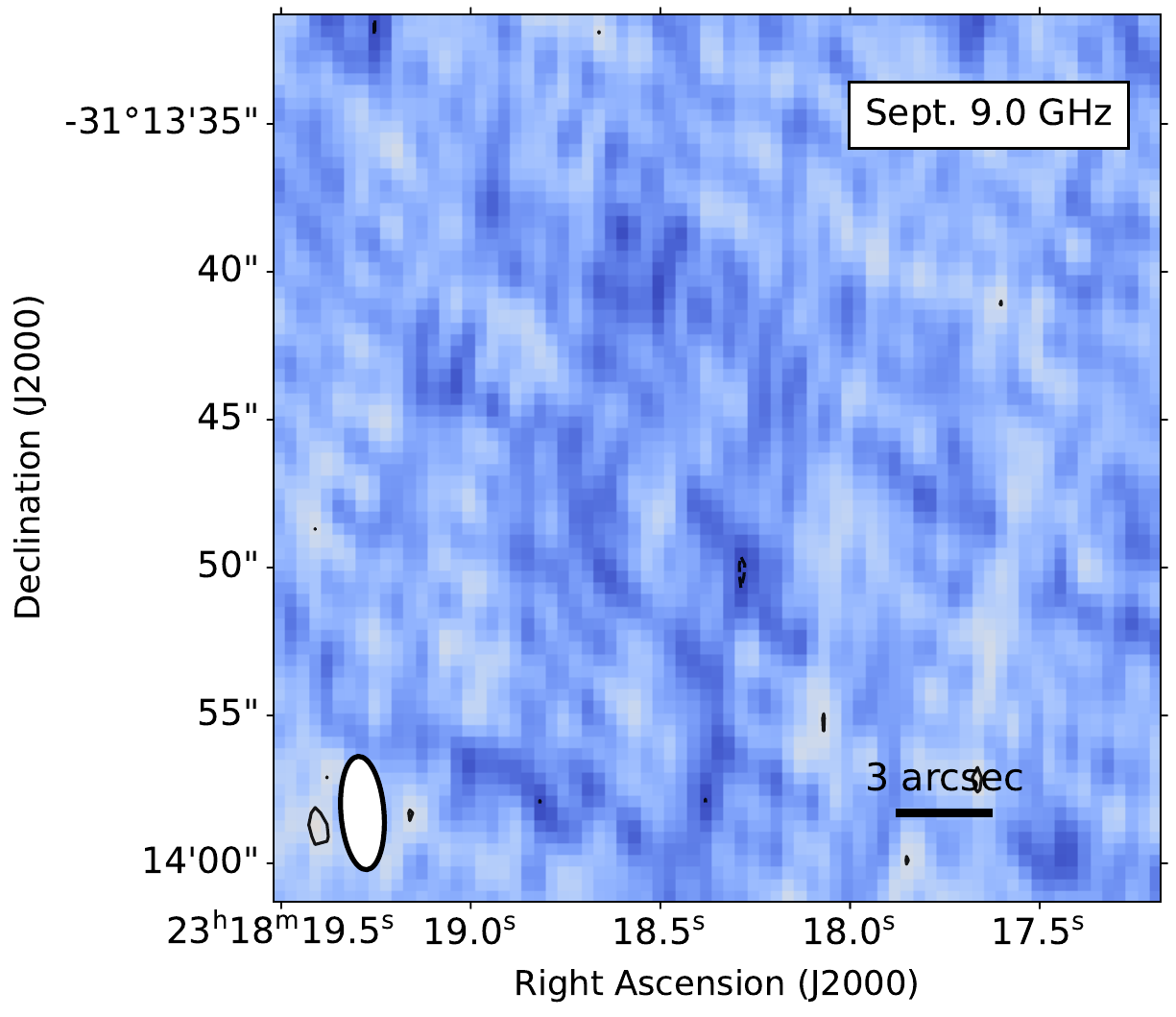}
    \caption{ATCA observation of VIK~J2318$-$31 at 2.1, 5.5 and 9.0~GHz (from left to right) in the two different epochs (April 2022, top, and September 2022, bottom). Contours start at $\pm$3$\times$RMS and increase by a factor of $\sqrt{2}$.}
    \label{fig:ATCA}
\end{figure*}

In this section we report the radio images obtained with the MWA (Fig. \ref{fig:MWA}), the uGMRT (Fig. \ref{fig:uGMRT}) and the ATCA (Fig. \ref{fig:ATCA}) telescopes of both epochs. A description of the reduction process is reported in Sec. \ref{sec:data}.

\end{document}